\documentclass{aa}  

\usepackage{graphicx}
\usepackage{array}    
\usepackage{amssymb}    
\usepackage{url}
\usepackage{color}
\usepackage{multirow}
\usepackage{makecell}

\usepackage{adjustbox}

\usepackage{txfonts}

\def\tef{$T_{\rm eff}$}

\begin{document} 

   \title{Periodic variable A-F spectral type stars in the northern TESS continuous viewing zone}

   \subtitle{I. Identification and classification}

   \author{M. Skarka
          \inst{1,2,3}
          \and
          J. \v{Z}\'{a}k\inst{4}
          \and
          M. Fedurco\inst{5}
          \and
          E. Paunzen\inst{2}
          \and
          Z. Henzl\inst{3,10}
          \and
          M. Ma\v{s}ek\inst{3,7}
          \and
          R. Karjalainen\inst{1}
          \and
          J.\,P. Sanchez Arias\inst{1}
          \and
          \hbox{\'{A}. S\'{o}dor\inst{6,9}}
          \and 
          R.\,F. Auer\inst{3}
          \and
          P. Kabáth\inst{1}
          \and
          M. Karjalainen\inst{1}
          \and
          J. Li\v{s}ka\inst{3,8}
          \and
          D. \v{S}tegner\inst{3}
          }
   \institute{Astronomical Institute of the Czech Academy of Sciences, Fri\v{c}ova 298, CZ-25165 Ond\v{r}ejov, Czech Republic\\
              \email{skarka@asu.cas.cz}
         \and
             Department of Theoretical Physics and Astrophysics, Masaryk University, Kotl\'{a}\v{r}sk\'{a} 2, CZ-61137 Brno, Czech Republic
        \and
            Variable Star and Exoplanet Section of the Czech Astronomical Society, Vset\'{i}nsk\'{a} 941/78, 757 01 Vala\v{s}sk\'{e}, Mezi\v{r}\'{i}\v{c}\'{i}, Czech Republic
        \and
            European Southern Observatory, Karl-Schwarzschild-str. 2, D-85748, Garching, Germany
        \and
            Department of Theoretical Physics and Astrophysics, Institute of Physics, Faculty of Science, University of Pavol Jozef \v{S}af\'{a}rik, Park Angelinum 9, SK-04154 Ko\v{s}sice, Slovakia
        \and
            Konkoly Observatory, Research Centre for Astronomy and Earth Sciences, Konkoly Thege Mikl\'{o}s  \'{u}t 15-17, H–1121 Budapest, Hungary
        \and
            FZU - Institute of Physics of the Czech Academy of Sciences, Na Slovance 1999/2, CZ-18221, Praha, Czech Republic
        \and
            Central European Institute of Technology, Brno University of Technology, Purk\v{n}ova 123, CZ-61200 Brno, Czech Republic
        \and
            MTA CSFK Lend\"ulet Near-Field Cosmology Research Group, Konkoly Thege Mikl\'{o}s  \'{u}t 15-17, H–1121 Budapest, Hungary
        \and
            Hv\v{e}zd\'{a}rna Jaroslava Trnky ve Slan\'{e}m, Nosa\v{c}ick\'{a} 1713, Slan\'{y} 1, CZ-27401 Slan\'{y}, Czech Republic
             }

   \date{Received May 16, 2022; accepted July 21, 2022}

 
  \abstract
   {In the time of large space surveys that provide tremendous amounts of precise data, it is highly desirable to have a commonly accepted methodology and system for the classification of variable stars. This is especially important for A-F stars, which can show intrinsic brightness variations due to both rotation and pulsations.}
   {The goal of our study is to provide a reliable classification of the variability of A-F stars brighter than 11\,mag located in the northern TESS continuous viewing zone. We also aim to provide a thorough discussion about issues in the classification related to data characteristics and the issues arising from the similar light-curve shape generated by different physical mechanisms.}
   {We used TESS long- and short-cadence photometric data and corresponding Fourier transform to classify the variability type of the stars. We also used spectroscopic observations to determine the projected rotational velocity of a few stars.}
   {We present a clear and concise classification system that is demonstrated on many examples. We find clear signs of variability in 3025 of 5923 studied stars (51\,\%). For 1813 of these 3025 stars, we provide a classification; the rest cannot be unambiguously classified. Of the classified stars, 64.5\,\% are pulsating stars of $g$-mode $\gamma$ Doradus (GDOR) and $p$-mode $\delta$\,Scuti types and their hybrids. We realised that the long- and short-cadence pre-search data conditioning simple aperture photometry data can differ significantly not only in amplitude but also in the content of instrumental and data-reduction artefacts, making the long-cadence data less reliable. We identified a new group of stars that show stable light curves and characteristic frequency spectrum patterns (8.5\,\% of the classified stars). According to the position in the Hertzsprung-Russell diagram, these stars are likely GDOR stars but are on average about 200\,K cooler than GDORs and have smaller amplitudes and longer periods. With the help of spectroscopic measurements of $v\sin i$, we show that the variability of stars with unresolved groups of peaks located close to the positions of the harmonics in their frequency spectra (16\,\% of the classified stars) can be caused by rotation rather than by pulsations. We show that without spectroscopic observations it can be impossible to unambiguously distinguish between ellipsoidal variability and rotational variability. We also applied our methodology to three previous studies and find significant discrepancies in the classification.}
   {We demonstrate how difficult the classification of variable A-F stars can be when using only photometric data, how the residual artefacts can produce false positives, and that some types cannot actually be distinguished without spectroscopic observations. Our analysis provides collections that can be used as training samples for automatic classification.}

   \keywords{Stars: variables: general --
                Stars: oscillations --
                Stars: rotation --
                Methods: data analysis --
                Catalogs
               }
    \maketitle
%

\section{Introduction}

The photometric space missions have provided us with invaluable insight into the mechanisms that produce variations in brightness. The ultra-precise data are especially important in the region of the Hertzsprung-Russell diagram where A-F stars are located ($6000<T_{\rm eff}<10000$\,K). This is the location where we can observe the transition between slow and fast rotation of stars, energy transfer via radiation and convection, and the transition between complex local magnetic fields and stable fossil fields. Stars in this region can also pulsate, both in acoustic ($p$) and gravity ($g$) modes.

We can thus observe several types of variability, often present at the same time. Among A-F pulsating stars, we can find: high-order $g$-mode $\gamma$ Doradus (GDOR) type pulsations with periods of the order of hours to days \citep{Balona1994,Kaye1999} generated by the convective-flux blocking mechanism \citep{Guzik2000,Dupret2005}; and $\delta$\,Scuti (DSCT) low-radial-order $p$-mode pulsators \citep{Breger2000} with pulsations excited by the opacity mechanism in the He\,\textsc{ii} layer \citep{Cox1963,Breger2000} and by turbulent pressure \citep{Houdek2000,Antoci2014} with periods of hours. There are also hybrid stars that show both DSCT and GDOR pulsations \citep{Henry2005,Handler2009,Grigahcene2010,Sanchez2017}.  SX Phe stars, which are Population II stars, have similar pulsational characteristics as DSCTs, which are Population I stars \citep{Balona2012}. Another type of pulsation appearing among chemically peculiar A-F stars with strong magnetic fields are p-mode oscillations with periods of the order of minutes observed in rapidly oscillating (roAp) stars \citep{Kurtz1982}. GDOR, DSCT, and roAp stars observed by the Transiting Exoplanet Survey Satellite \citep[TESS,][]{Ricker2015} in sectors 1 and 2 were thoroughly studied by \citet{Antoci2019} and \citet{Cunha2019}. 

Rotation can also cause photometric variations. In stars with stable atmospheres, light elements (e.g. He) can diffuse downwards, and others (usually Si and rare-earth elements, such as Sr, Cr, and Eu) can be levitated to the surface and create chemical spots \citep{Michaud1970, Preston1974}. In the presence of a magnetic field, these spots can be sustained and cause rotational modulation \citep{Stibbs1950,Kochukhov2011}. In some of the A-F type stars, variability similar to solar activity and spots was observed \citep{Balona2013}. Rapidly oscillating Ap stars also often show rotational modulation. Those observed in the first two TESS sectors were studied by \citet{Cunha2019}.

Among rotationally variable stars, we can also find ellipsoidal variables, that is, stars in non-eclipsing binary systems that are tidally deformed and, due to gravitational darkening, can also show brightness variations as the stars orbit around the common centre of mass \citep{Morris1985,Beech1985}. For completeness, we should not forget to mention classical variable stars such as RR Lyrae stars \citep{Catelan2015} and systems containing the intrinsic variable stars mentioned above, such as eclipsing binaries and exoplanetary candidates. Flare-like events have also been observed in A-type stars \citep{Balona2012b}. However, \citet{Pedersen2017} find that the flares can come from nearby stars and companions in binary systems, raising doubts about flares among A-type stars.

The classification of variable stars observed by space missions is usually based on common light-curve characteristics and on characteristics of the frequency spectra \citep[e.g.][]{Debosscher2011,Uytterhoeven2011,Balona2011gdor,Balona2011rot,Bradley2015}. Most of the classifications have been performed semi-automatically employing visual inspection. Nowadays, with the tremendous increase in data quantity, automatic procedures using (supervised) machine learning and neural networks are being developed \citep[e.g.][]{Debosscher2011,Audenaert2021}. Although there are some basic common criteria, the classification methodology is not unified and the classifications can differ. In addition, due to the data quality, data characteristics, and/or similar manifestations of different physical mechanisms, ambiguities in classification can emerge. Therefore, it is important to properly identify the physical mechanism causing the light variations and classify the stars accordingly. 

In this paper we investigate A-F stars observed by the TESS space mission \citep{Ricker2015} near the northern ecliptic pole (Sects. \ref{Sect:TESS} and \ref{Sect:Sample}). The time base of the TESS observations of the stars in this region is the longest available, providing a great opportunity for the most reliable classification of periodically variable stars and studying changes with periods of the order of days to tens of days. We perform a careful case-by-case visual classification of the variability type of every particular star using 30- and 2-minute cadence data  and provide a large catalogue of variable stars. The motivation for this work was to identify interesting targets for detailed study. However, a deeper investigation of particularly interesting targets is a topic for separate paper(s). 

A significant portion of the paper is dedicated to the discussion of effects that influence the identification of variability, false variability connected with the instrumental and data-reduction artefacts, and how the data themselves affect the results (Sects.~\ref{Sect:Identification}, \ref{Sect:Classification}, and \ref{Sect:AmbiguityClassification}). We describe the precise criteria adopted for the classification and discuss ambiguities across different variability types (Sect.~\ref{Sect:AmbiguityClassification}). The results of our analysis are given in Sect.~\ref{Sect:Discussion}. The summary of the paper and future prospects are in Sect.~\ref{Sect:Conclusions}.

\section{Observations and data products}\label{Sect:TESS}

The heart of the TESS satellite consists of four cameras with four 2K$\times$2K CCDs each that produce combined field of view (FOV) of 24$^{\circ}\times96^{\circ}$ \citep{Ricker2015}. Due to the large FOV, the angular resolution per pixel is only 21"/px. The part of the sky that is observed for two consecutive TESS orbits ($\sim 2\times 13.7$ days) around the Earth is called a 'sector'. After the observation of a sector is finished, the FOV moves 27$^{\circ}$ along the ecliptic for the next sector. The whole hemisphere is scanned in a 'cycle'. TESS avoids regions closer than 6$^{\circ}$ to the ecliptic plane to eliminate the disruptive light of the Solar System bodies and the Moon. This observing strategy means that the sectors overlap around the ecliptic poles, the TESS continuous viewing zone (CVZ) providing almost uninterrupted observations with the time base of about 350\,days. Such data are the most suitable for investigations of stellar variability.

The full frame images (FFIs) are downloaded every 30 minutes and serve as an input for a so-called long-cadence (LC) photometry that is provided for all observed objects down to 16-17\,mag. Dedicated postage-stamp images, target pixel files (TPFs), of selected targets are downloaded every 2 minutes. These data serve as an input for a so-called short-cadence (SC) photometry. The TPFs and the light curves processed by the TESS Science Processing Operations Center \citep[SPOC;][]{Jenkins2016} are available at the Mikulski Archive for Space Telescopes (MAST)\footnote{https://archive.stsci.edu/}. SPOC provides two kinds of photometric data -- simple aperture photometry (SAP) and pre-search data conditioning (PDC) SAP flux with long-term trends removed \citep[][]{Twicken2010}. At MAST, also products from the quick-look pipeline \citep[QLP;][]{Huang2020a,Huang2020b} with the 30-min sampling are available.

\subsection{Sample selection and data retrieval}\label{Sect:Sample}

To select stars that are close to the TESS CVZ, we first cross-matched the TESS Input Catalogue (TIC) v8.0 \citep{Stassun2019} with the SIMBAD database \citep{Wenger2000} around the northern ecliptic pole (RA=18$^{\rm h}$00$^{\rm m}$00$^{\rm s}$, Dec.=+66$^{\circ}$30'00'') with the radius of 15\,deg using the CDS X-Match service \citep{Boch2012,Pineau2020}. The cross-match with the SIMBAD database gives a good chance that there is some available information about the stars in the literature if needed. The initial sample for our work consists of 67093 stars in and close to the TESS CVZ.

In the next step, we limited our sample to stars with 6\,000$<$\tef$<$10\,000\,K to get only stars with A-F spectral types\footnote{We used \tef~ from the TIC catalogue \citep{Stassun2019}}. We also limited our sample to stars with brighter than 11\,mag in Johnson $V$ filter to select only stars with good quality data that are bright enough to be suitable for spectroscopic observations with 1-metre class telescopes and for further studies. After omitting duplicates, 5923 stars were accepted for the analysis. The basic distribution of the temperatures is shown in the upper part of Table~\ref{Tab:TefMags}, while the brightness distribution of the sample stars is shown in the bottom part of Table~\ref{Tab:TefMags}.

\begin{table}
\caption{Distribution of the number of stars in particular temperature (top part) and brightness (bottom part) ranges of the sample stars \citep[from the TIC;][]{Stassun2019}.}             
\label{Tab:TefMags}      
\centering                          
\begin{tabular}{c c}        
\hline\hline                 
$T_{\rm eff}$ (K) & N \\    
\hline                        
$6000-7000$ & 4305 \\      
$7000-8000$ & 1110 \\ 
$8000-9000$ & 353 \\ 
$9000-10000$ & 155 \\  \hline \hline
$V$ (mag) & N \\ \hline
$<6$ & 31 \\
$6-8$ & 229 \\
$8-10$ & 2017 \\
$>10$ & 3646 \\
\hline                                   
\end{tabular}
\end{table}

Since our work started in February of 2021, we only downloaded data observed in Cycle 2, and thus we only analysed data from sectors 14-26 (July 2019-June 2020). SPOC SC and LC data (3403 and 5787 stars, respectively) and QLP LC data (5923 stars) were obtained from the MAST archive using the \textsc{Lightkurve v2.0} software \citep{Lightkurve2018,Barentsen2020}. \textsc{Lightkurve v2.0} was also used for merging the data from different sectors. The median time span of the datasets is 352 days for both LC and SC SPOC data, respectively. This means that the vast majority of the studied stars were observed in sectors spread out over the whole of cycle~2. The median number of points per dataset is 12000 and 170000 for LC and SC data, respectively. 

\begin{figure}
\centering
\includegraphics[width=0.47\textwidth]{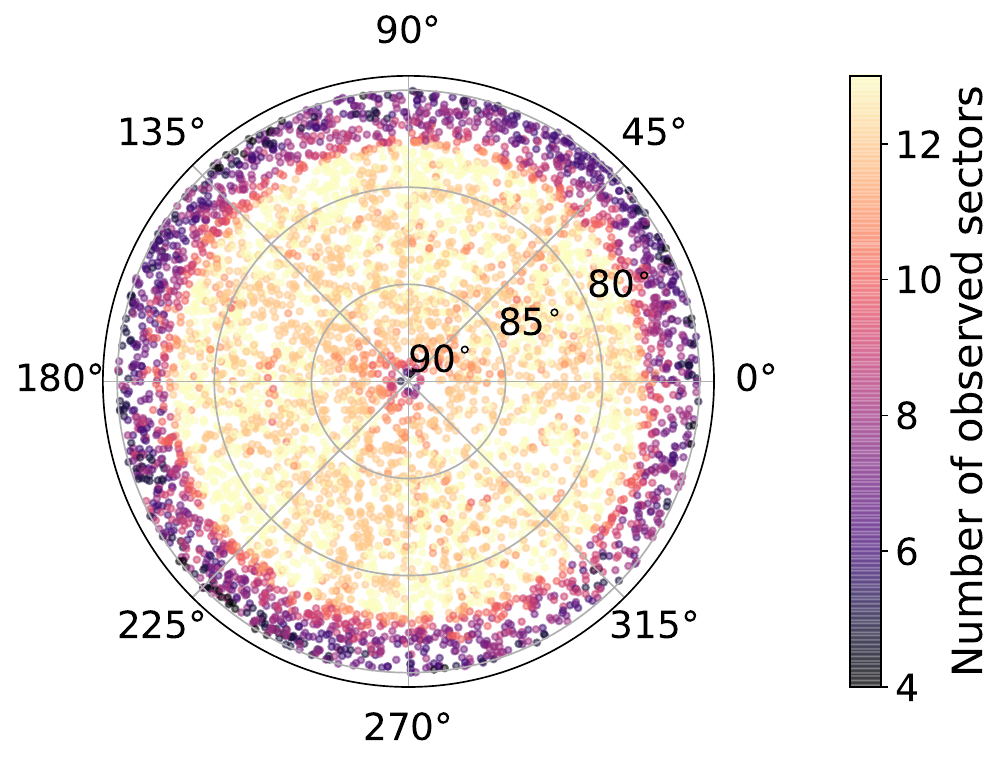}

\includegraphics[width=0.47\textwidth]{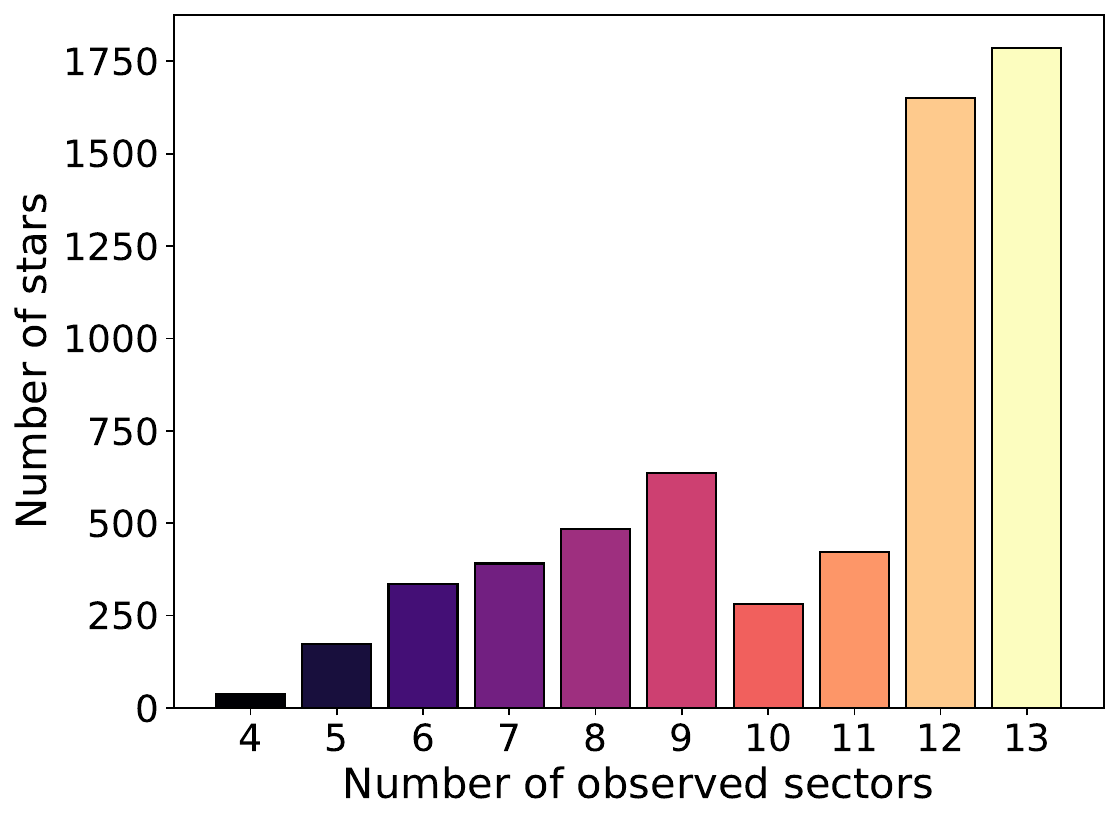}
\caption{Position of the investigated stars around the northern ecliptic pole in the ecliptic coordinates (top panel). The colour coding shows the number of sectors in which the data were available in the SPOC routine. The bottom panel shows in how many sectors the data were available for how many stars.}
\label{Fig:Hist_sectors}
\end{figure}

The spatial distribution and number of sectors in which the particular stars were observed are shown in Fig.~\ref{Fig:Hist_sectors}. We note here that the SPOC products were not always available for all the sectors in which the star was observed. For example, there should be data for TIC~229412873 available from 12 sectors from Cycle 2, but there are SPOC data products only for three sectors (all 12 sectors in the QLP routine). 

For the forthcoming analysis, we transformed the immediate flux $F_{i}$ to relative magnitudes $\Delta m_{i}$ by using Pogson's equation:
\begin{equation}
    \Delta m_{i} = -2.5\log \left( \frac{F_{i}}{F_{\rm mean}} \right),
\end{equation}
where $F_{\rm mean}$ is the mean flux of the observations.

We have not applied any additional cleaning or detrending procedures, filtering of the data or outlier removal. We simply used the PDCSAP data as they are provided, because careful examination of selected data, show they are almost free of strong outlier points or glitches. The additional modification of the datasets would need individual approaches for each of the stars in the sample, which is beyond the scope of this paper.

\subsection{Spectroscopic supporting data}\label{Subsect:Spectroscopy}

To help to better classify the stars, we gathered available low-resolution spectra of 126 of our targets from the Large Sky Area Multi-Object Fiber Spectroscopic Telescope (LAMOST; $R=\lambda/ \Delta \lambda \approx 1800$, spectral range $3700-9000$\,\AA) of the Chinese Academy of Science \citep{Cui2012,Zhao2012} and determined their spectral type by using the {\textsc MKCLASS} code \citep{Gray2014}. This software is able to recover the spectral type with results that are comparable to the manual classification \citep[e.g.][]{Gray2014}.  

We also gathered new high-resolution spectra ($R=\lambda/ \Delta \lambda \approx 40000$ at H$\alpha$, spectral range $3800-9100$\,\AA) to prove the rotational variability via projected rotational velocity of six stars that may be classified as pulsating stars. The stars were observed with the Ond\v{r}ejov Echelle Spectrograph (OES) mounted on the 2m Perek Telescope, Academy of Sciences, Czech Republic \citep{Kabath2020}. All the spectra had signal-to-noise ratio ({\it S/N}) above 100 near H$\alpha$.

The spectra were modelled with the {\textsc iSpec} software \citep{Blanco-Cuaresma2014,Blanco-Cuaresma2019}. As the input parameters for the fitting with radial transfer code {\textsc SPECTRUM} \citep{Gray1994} and MARCS model atmosphere \citep{Gustafsson2008}, we used the values of \tef~and $\log g$~from the TIC catalogue \citep{Stassun2019} and adjusted them to better fit the spectra if necessary. Subsequently, we fixed these values and left $v \sin i$ as a free parameter and determined the projected rotational velocity.

\section{Identification of variability and data shortcomings}\label{Sect:Identification}

Stellar variability in the space data is usually investigated in the time (light curve) and frequency domains via Fourier transform (FT) of the data \citep[e.g.][]{Balona2011rot,Bradley2015}. We created the FT of the LC and SC SPOC and QLP data\footnote{In the range of 0-300 and 0-24\,c/d for the SC and LC data, respectively.} by using a Python implementation of the non-uniform fast FT\footnote{\url{https://github.com/dfm/python-nufft}} that is significantly faster than a classical Lomb-Scargle periodogram \citep{Lomb1976,Scargle1982} but gives basically the same results for TESS data.

The periodic signal with frequency $f$ in the data is usually considered as real if the {\it S/N} of the corresponding peak in the frequency spectrum is larger than four \citep{Breger1993}. However, higher {\it S/N} might be needed in case of stars with dense Fourier peaks \citep{Bognar2020}. We define the {\it S/N} as the ratio of the amplitude of the peak at a given frequency $f$ and the mean amplitude of the peaks in the vicinity of the peak $f\pm1$\,c/d. This is a more conservative approach than using power-amplitude spectra or defining the {\it S/N} as the ratio of the amplitude of the peak and the standard deviation of the peaks in the immediate vicinity, as is often done \citep[e.g.][]{Balona2011rot,DeMedeiros2013}. We did not apply any amplitude threshold \citep[as was done e.g. by][]{Uytterhoeven2011,Bradley2015} because each target has a different noise level.  

\begin{figure}
\centering
\includegraphics[width=0.47\textwidth]{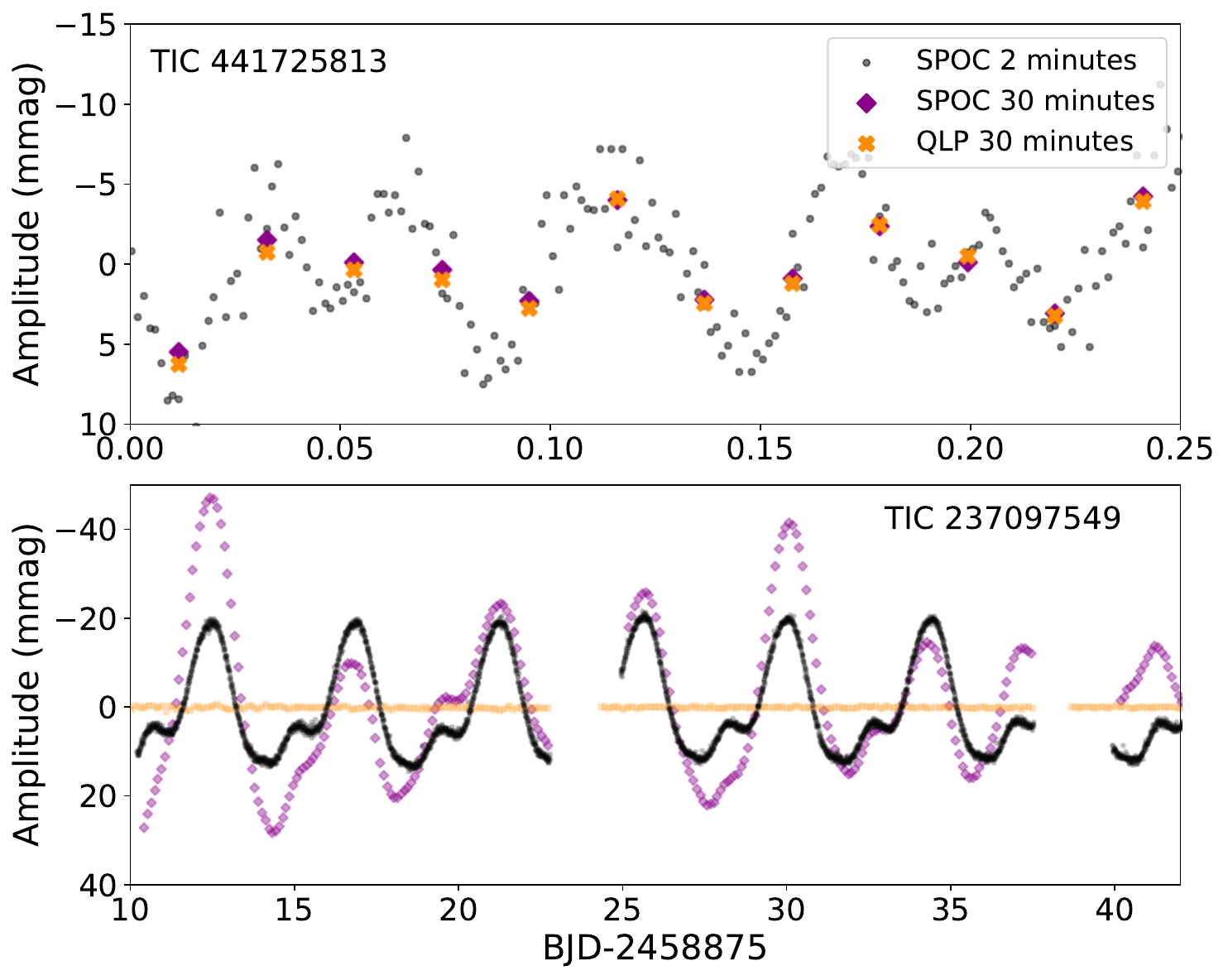}
\caption{Comparison of the data with different samplings obtained using different routines. In low-amplitude short-period variable stars, the LC data from QLP and SPOC data products give similar results but with smaller amplitudes (the upper panel). The bottom panel shows that in stars with amplitudes of the order of tens of millimagnitudes, the LC data can already suffer from some instrumental variations (magenta points), but the QLP routine (orange points) fully removes the variability. In both stars, the LC data are less reliable than the SC data in determining the amplitudes of brightness variations and in determining the variability type.}
\label{Fig:Comparison}
\end{figure}

\begin{figure}
\centering
\includegraphics[width=0.47\textwidth]{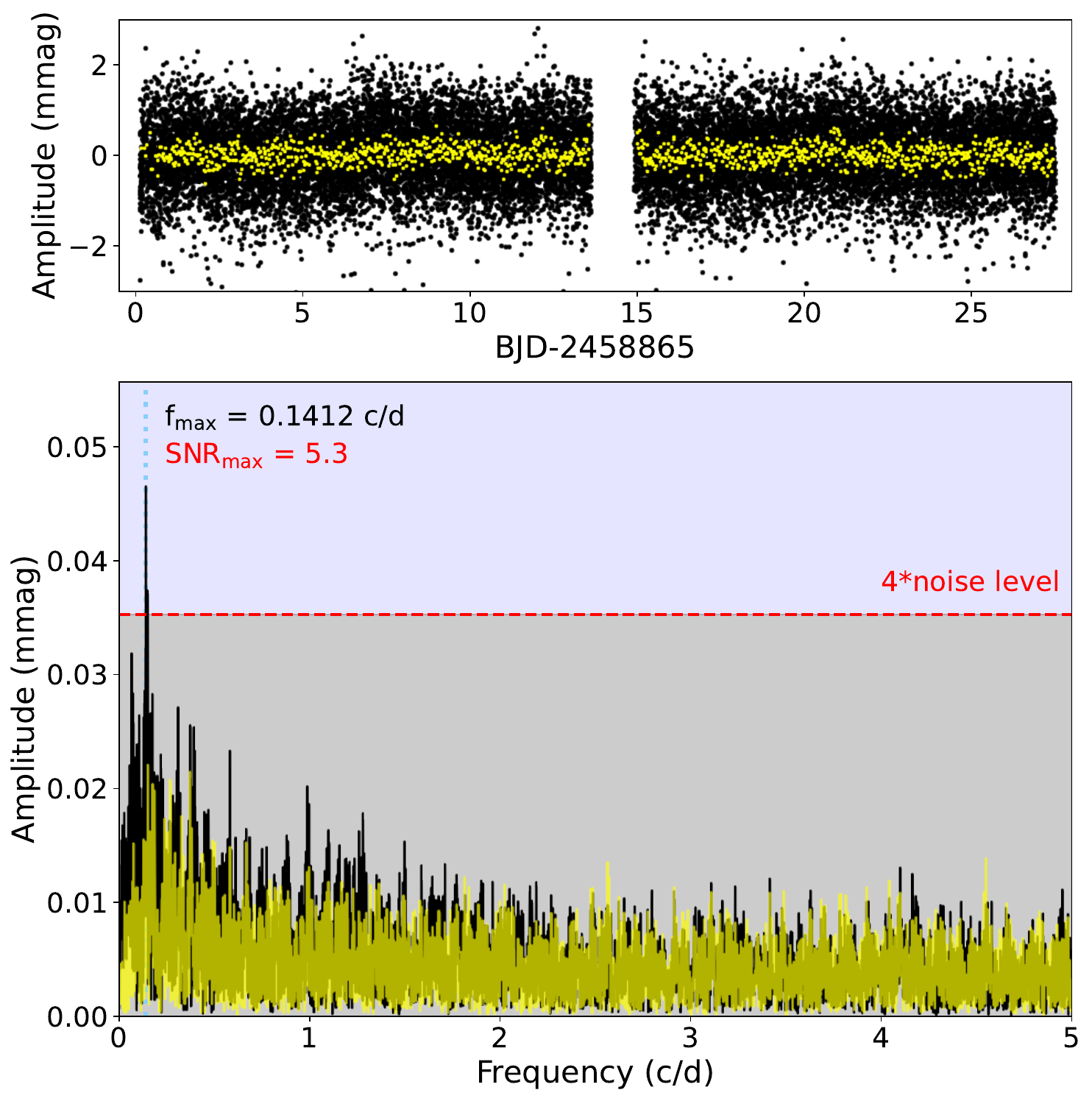}
\caption{SPOC data of TIC~2116199 from sector 14 (top panel) and frequency spectra based on four available sectors with a time span of 352\,days (bottom panel). The LC data and the corresponding frequency spectrum are shown with yellow, and the SC data and the corresponding frequency spectrum with black. The horizontal dashed red line shows four times the noise level limit, while the vertical dotted blue line shows the most significant frequency.}
\label{Fig:SNR}
\end{figure}

The basic problem in the identification (and classification of the variability in general) arises from the natural characteristics of the data (noise, sampling, time span of the data, data reduction and detrending) in combination with the variability timescales and amplitudes. We illustrate these issues in Figs.~\ref{Fig:Comparison} and \ref{Fig:SNR}. The data reduced with different routines and with different sampling must naturally be different. As is apparent from the top panel of Fig.~\ref{Fig:Comparison}, the amplitude of the light variations in the LC data have smaller amplitudes averaging out the fast variations. 

The QLP routine removes the large-amplitude variations by applying a high-pass filter \citep[the bottom panel of Fig.~\ref{Fig:Comparison}, orange points,][]{Huang2020a}. From the bottom panel of Fig.~\ref{Fig:Comparison}, it is also apparent that residual instrumental and reduction artefacts due to imperfect background subtraction and/or poorly defined aperture might be present in the data (magenta diamonds, SPOC LC data). These artefacts appear more often in the SPOC LC data than in the SPOC SC or QLP data and can cause serious issues when classifying variability types. In the worst case, the artefacts can lead to a completely wrong classification. The QLP routine can even fully remove the variations (Fig.~\ref{Fig:Comparison}, bottom panel). A nice comparison of the available data products and their suitability for the large-amplitude variables investigation, particularly for Cepheids and RR Lyrae stars, observed by the Kepler \citep{Borucki2010} and TESS missions, can be found in \citet{Plachy2019} and \citet{Molnar2022}, respectively.

Figure~\ref{Fig:SNR} shows an example of how the frequency spectra for the SPOC data with different sampling can differ. There is a significant peak with {\it S/N}\ $>4$ in the SC data (black line and points), while there is no significant peak in the FT of the LC data (yellow line and points). According to the SC data and the peak, TIC~2116199 should be considered as a variable star. However, because the FT of LC data shows no significant peak and there is actually no apparent variability in the light curve, we assume that the possible variation is of artificial nature and classify TIC~2116199 as a non-variable star.

From all the above mentioned reasons, we performed careful individual investigation of all available datasets for each star giving the preference to the SPOC SC data (where available) that seem to be the most reliable. The identification of variability and subsequent classification was performed on an individual basis. In addition, in case of LC data, the cadence is a significant fraction of the DSCT period and longer than that of a roAp star. Further shortcomings of the classification are discussed in Sect.~\ref{Sect:AmbiguityClassification}.

\section{Classification scheme and notation}\label{Sect:Classification}

Similar to the identification of real variations (see Fig.~\ref{Fig:Comparison} and \ref{Fig:SNR}), classification based solely on the photometric data can be a difficult task, leading to an ambiguous solutions. Usually, multi-parametric classification employing periods, light curves, and frequency spectra characteristics are used \citep[e.g.][]{Uytterhoeven2011,Balona2011gdor,Bradley2015}. Some types of variable stars are recognizable upon a first look (typically the high-amplitude variables, for instance, eclipsing binaries of Algol type), but in some variability types the classification may even be impossible. For example, the light curve of an ellipsoidal variable can look the same (including periods and amplitudes) as an over-contact eclipsing variable of W UMa type, pulsating RRC star, or spotted chemically peculiar star; pulsations can mimic rotation, eclipses, and so on. (see Sect.~\ref{Sect:AmbiguityClassification}). 

\begin{figure*}
\centering
\includegraphics[width=0.98\textwidth]{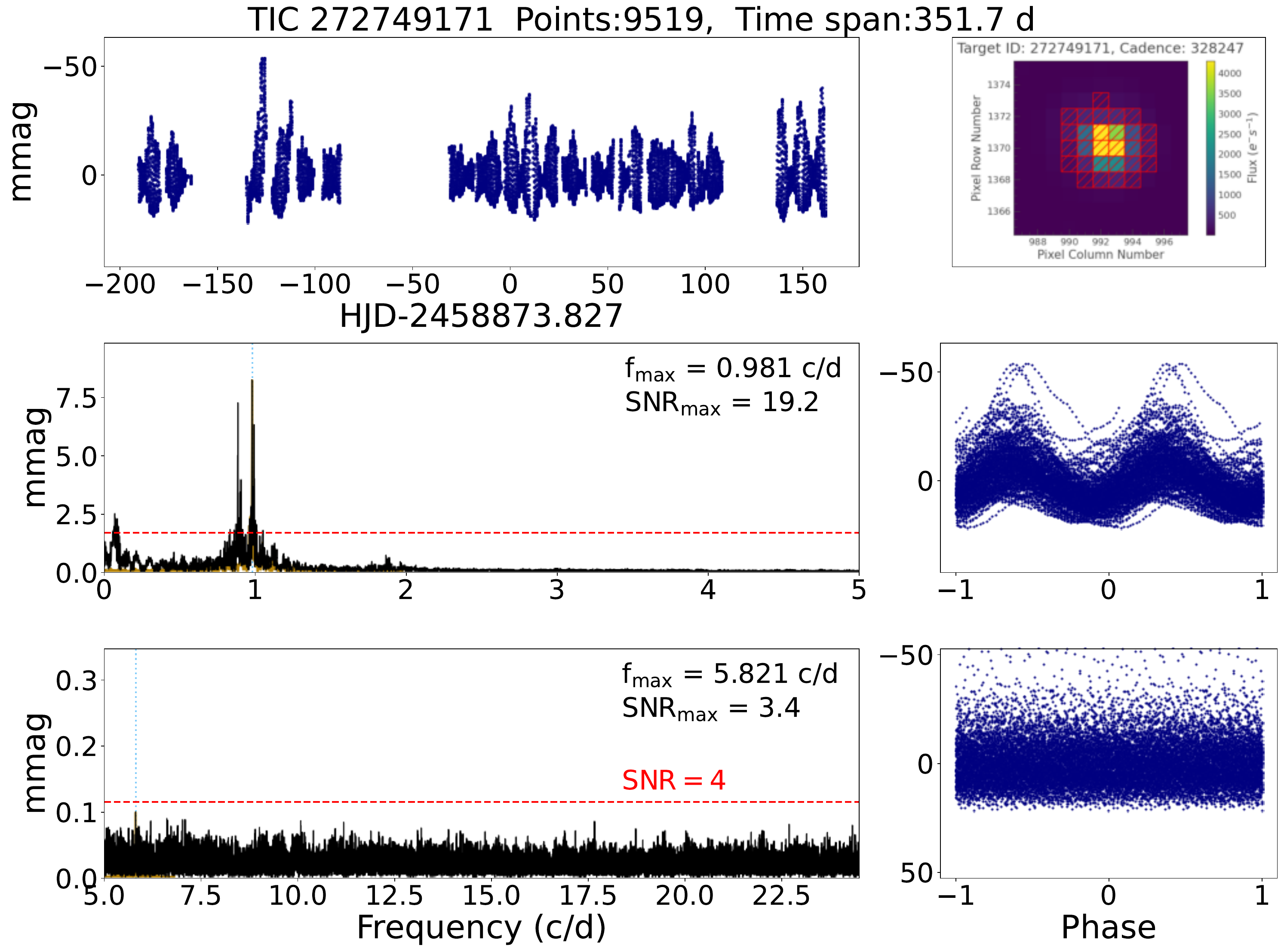}
\caption{Example of the figures that have been produced and used for the basic classification of each of the stars in the sample. The upper left panel shows the data, the top right panel shows the TPF with the aperture, and the middle and bottom panels show the frequency spectra (left) and the data phase-folded with the most prominent peaks in the frequency spectra (right).}
\label{Fig:Overview}
\end{figure*}

For each star, we produced a figure showing the time series data, data phase-folded with the dominant frequencies below and above 5\,c/d, a TPF with the aperture, and frequency spectra divided into parts below and above 5\,c/d (Fig.~\ref{Fig:Overview}). This frequency limit is a usually accepted value dividing the GDOR and DSCT pulsations \citep{Grigahcene2010} that has been applied in many studies dealing with the data obtained from space \citep[e.g.][]{Uytterhoeven2011,Balona2011gdor,Bradley2015}. Plots shown in Fig.~\ref{Fig:Overview} served as the main decision factors for the classification. If needed, particular stars were investigated more closely.

The basic variability types were adopted from the Variable Stars Index (VSX) catalogue\footnote{\url{https://www.aavso.org/vsx/index.php?view=about.vartypes}} \citep[][]{Watson2006}. The notation is based on the classification system by the General Catalogue of Variable Stars \citep[GCVS;][]{Samus2009}. For the stars that are certainly variable but we were not able to determine the variability type, we used 'VAR'. If there is ambiguity between two or more variability types, we use the pipe symbol '|' meaning 'or'. If a star shows a combination of variability types, we give plus sign with the variability type of larger amplitude listed first. 

Because the temperature of the stars in our sample covers a wide range, the expected mixture of physical mechanisms of variability are also very broad. Thus, we examined the general characteristics of the light variations and put an emphasis on explaining the physical mechanism of the variability. 

We considered finer divisions to more specific types only in high-amplitude stars (for example, EA versus EB, RRAB versus RRC, etc.). There were attempts to perform a more delicate classification also in low-amplitude classes of variables. For example, \citet{Balona2011gdor} introduced three subclasses of GDOR stars according to their light-curve shape: symmetric (SYM), asymmetric (ASYM), and multiple modes (MULT). Similarly, \citet{Balona2011rot} introduced three classes of rotationally variable stars: SPOT (multi-periodic with dominant frequency), SPOTM (clear travelling wave), and SPOTV (clear dominant period). Similar notation as for GDOR stars created by \citet{Balona2011gdor} was introduced also for DSCT stars by \citet{Bradley2015}. However, we did not perform such a detailed classification. The reasons are that there can be ambiguity in the variability types. Instead, we decided to build a classification on the assumed physical mechanisms rather than on the look of the light curve (see Sect.~\ref{Sect:Classification} and \ref{Sect:AmbiguityClassification}). We also did not distinguish between DSCT and SX Phe stars and do not divide DSCT into LADS and HADS classes \citep{Frolov1984,Petersen1996}. Similarly, we do not distinguish between low- and high-amplitude GDOR stars \citep{Paunzen2020}.

The variability types and criteria for classification are listed in Tables~\ref{Tab:VarTypes_Bins}-\ref{Tab:VarTypes_ROT} including light curves and corresponding frequency spectra of the typical examples. The types and results are explored in details in Sections \ref{Sect:AmbiguityClassification} and \ref{Sect:Discussion}. In Sect.~\ref{Sect:AmbiguityClassification}, we discuss details and problems in classification. In Tables~\ref{Tab:VarTypes_Bins}-\ref{Tab:VarTypes_ROT}, we do not show the minor variability types identified in our sample. We identified two RRAB/BL stars (RR Lyrae of AB type showing the Blazhko effect)\footnote{Saw-tooth light-curve shape, harmonics of the basic frequency in FT} and one RRC star\footnote{More sinusoidal shape, harmonics of the basic frequency in FT} accompanied with one RRAB and two RRC candidates that have amplitudes only in the mmag range, which is very suspicious. These two RRC candidates may belong to a new class of horizontal-branch variables \citep{Wallace2019}. We also found five candidates of roAp stars that show frequencies above 100\,c/d.

In Table~\ref{Tab:Main}, we provide the identification, classification, frequency of the dominant peak in the FT and the amplitude of the brightness variation. We also give a cross-match of our sample with the VSX catalogue with the distance limit of 20". There are 172 variable stars from our sample in the VSX (version 2022-02-21), while in 10 of the VSX stars we did not find any variation. For eclipsing binaries, we also give the zero epoch $M_{0}$ of the primary eclipse in Table~\ref{Tab:Main}.

The amplitude given in Table~\ref{Tab:Main} is a rough visual estimate of the maximum amplitude (from minimum to maximum light) seen in the data. For example, the amplitude of TIC 272749171 shown in Fig.~\ref{Fig:Overview} is 70\,mmag. This value is only indicative and can be different from the real amplitude of the variations.
For example, the data might not contain the extrema of the variations, the light of the object can be contaminated by the light of nearby stars, the amplitude may be decreased by the LC in case of fast variations.

%
\renewcommand{\arraystretch}{1.2}
\newcolumntype{C}{>{\centering\arraybackslash\footnotesize}m{0.25\textwidth}}
\setlength{\tabcolsep}{4pt}
\begin{table*}[htbp!]
\caption{Typical examples of the variability types caused by the binarity of the stars. The notation is in the first column, 'Type', a description of the characteristic features present in the light curve is in the second column, and characteristic features in the FT are described in the third column. The assumed physical origin of the brightness variations is in the fourth column. For each variability type we show the parts of the light curves that reflect the basic variability for two example stars, including the FTs of the full available datasets. The Julian date is arbitrarily shifted. Note the different scales of the vertical axes.}             
\label{Tab:VarTypes_Bins}      
\centering     
\begin{adjustbox}{width=\textwidth,center}
\begin{tabular}{>{\centering}m{0.25\textwidth}CCC}
 
\hline\hline       
                     
Type & {\normalsize  Light curve} & {\normalsize Frequency spectrum} & \normalsize Physical origin \\ 
\hline                    
{\Large EA; EP} & 
constant light in maximum, sharp minima (with the same depth in the event of EP) & many well-defined harmonics of the basic frequency & 
eclipses in binary system with detached components; transits of an exoplanet\\ 
\multicolumn{4}{c}{\includegraphics[width=0.96\textwidth]{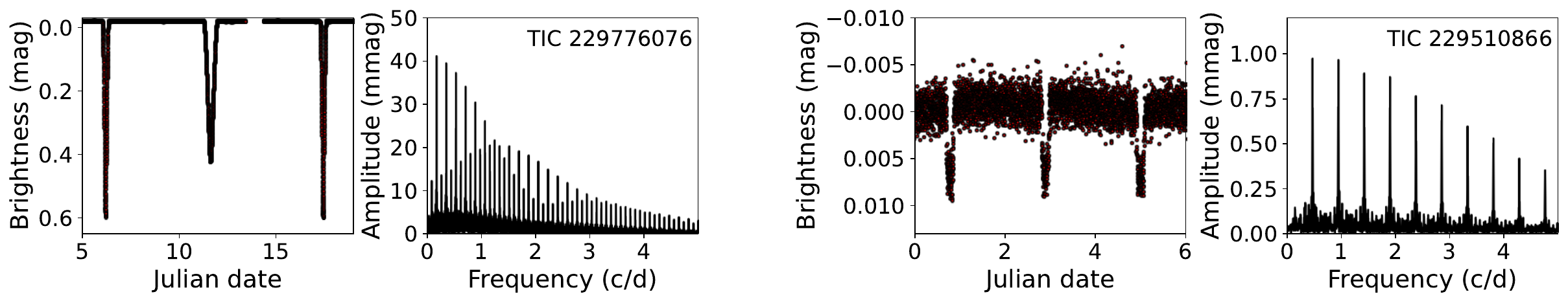}} \\ \hline

{\Large EB} & variation in maximum light, well-defined minima with (generally) different depths & many well-defined harmonics of the basic frequency with decreasing amplitude & eclipses in binary system with deformed component(s) \\
\multicolumn{4}{c}{\includegraphics[width=0.96\textwidth]{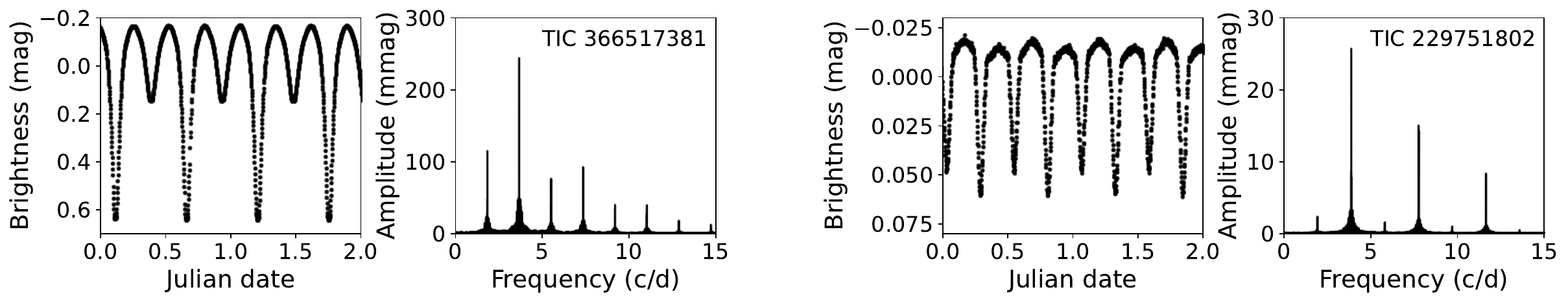}} \\ \hline

{\Large EW} & 
smooth brightness variation, flat bottom of eclipse(s) may be present, amplitudes $>100$\,mmag & 
well-defined harmonics of the basic frequency with quickly decreasing amplitudes towards higher frequencies & 
eclipses in binary system with components filling their Roche lobes, common envelope binaries \\
\multicolumn{4}{c}{\includegraphics[width=0.96\textwidth]{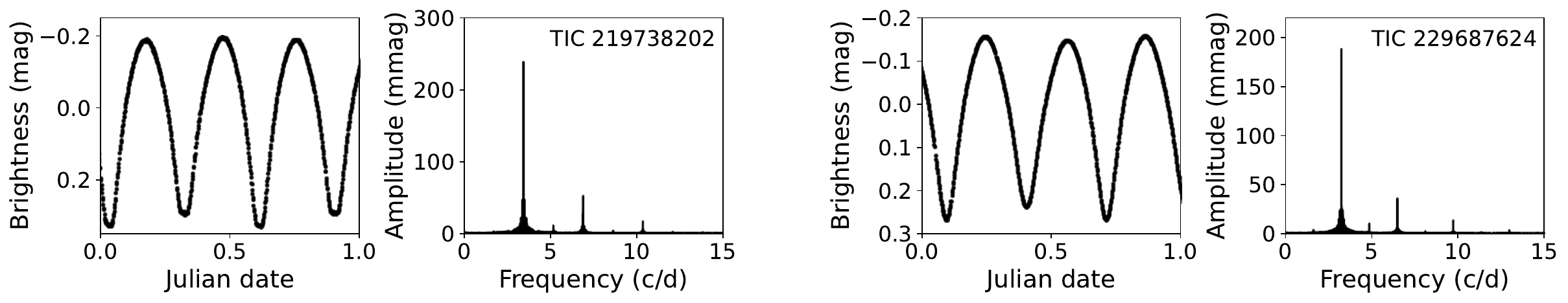}} \\ \hline

{\Large ELL} & 
smooth brightness variation, strictly symmetric and repeating light curve without sharp features, maxima have the same brightness & 
one or two dominant peaks at harmonics ($f_{1}=2f_{2}$), additional low-amplitude harmonics may be present, amplitudes $<100$\,mmag & 
non-eclipsing binary system with tidally deformed components  \\
\multicolumn{4}{c}{\includegraphics[width=0.96\textwidth]{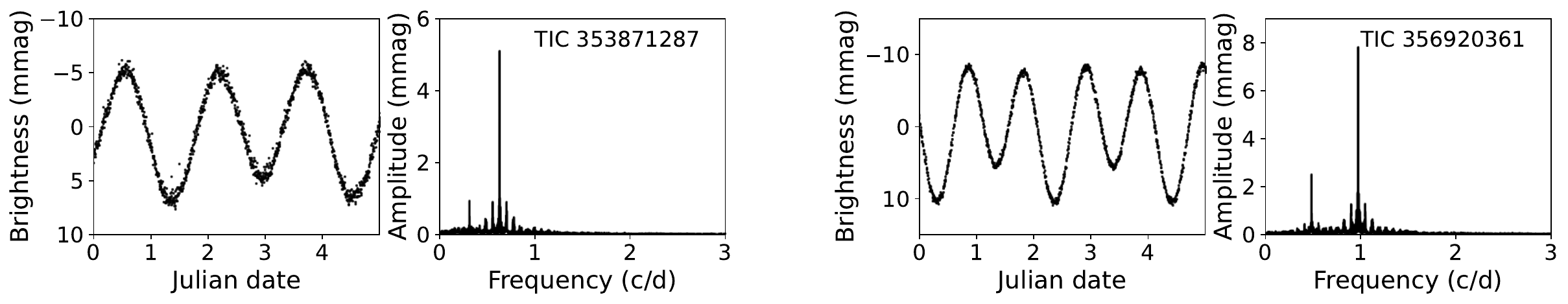}} \\ 

\hline                  
\end{tabular}
\end{adjustbox}
\end{table*}

%
\renewcommand{\arraystretch}{1.2}
\setlength{\tabcolsep}{4pt}
\begin{table*}[htbp!]
\caption{Typical examples of the variability types assumed to be caused by stellar pulsations. The columns are the same as in Table~\ref{Tab:VarTypes_Bins}.}             
\label{Tab:VarTypes_Puls}      
\centering  
\begin{adjustbox}{width=\textwidth,center}
\begin{tabular}{>{\centering}m{0.25\textwidth}CCC}
 
\hline\hline       
                     
Type & 
{\normalsize  Light curve} & 
{\normalsize Frequency spectrum} & 
\normalsize Physical origin \\ 

{\Large VAR} & 
weak or no signs of variability (low amplitude) and/or ambiguity in classification & 
single peak (with frequency usually below 1\,c/d) and/or peaks with unclear nature, peaks with {\it S/N} slightly above 4& 
unknown origin, possible stellar activity or instrumental artefacts \\ 
\multicolumn{4}{c}{\includegraphics[width=1.0\textwidth]{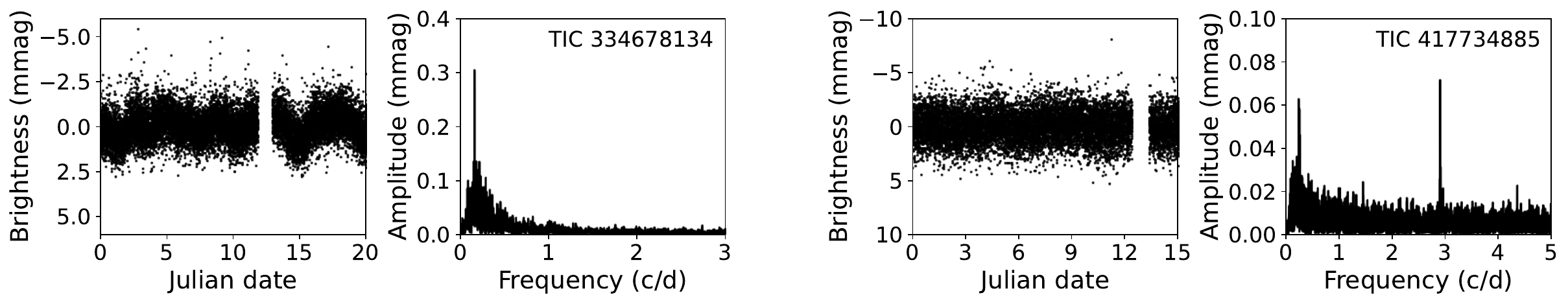}} \\ \hline

{\Large DSCT} & 
(ir)regular fast variations, beating, bumps, interference & 
two or more independent peaks above 5\,c/d & 
p-mode pulsations \\
\multicolumn{4}{c}{\includegraphics[width=1.0\textwidth]{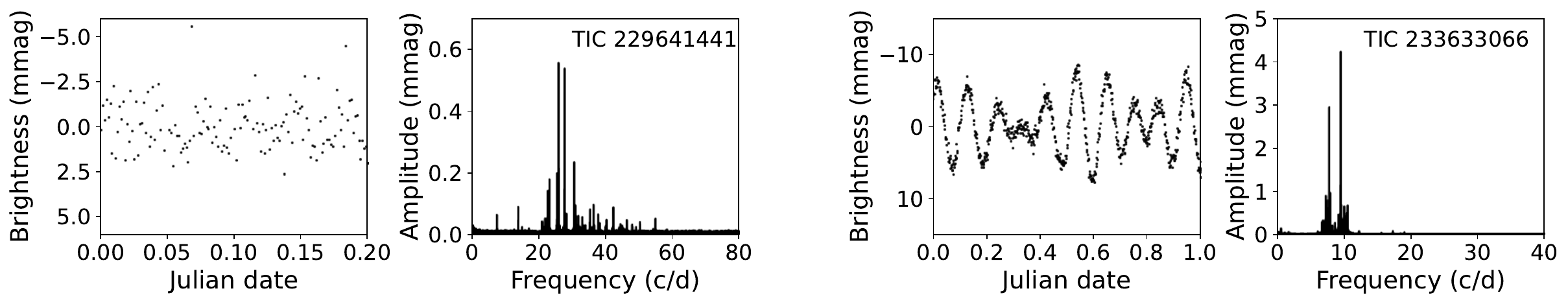}} \\ \hline

{\Large GDOR} & 
(ir)regular variations, beating, bumps, interference, sharp variations  & 
two or more independent peaks below 5\,c/d, peaks are usually in groups, well-defined single peaks are not harmonics, groups of peaks can be close to the positions of their harmonics & 
g-mode pulsations \\
\multicolumn{4}{c}{\includegraphics[width=1.0\textwidth]{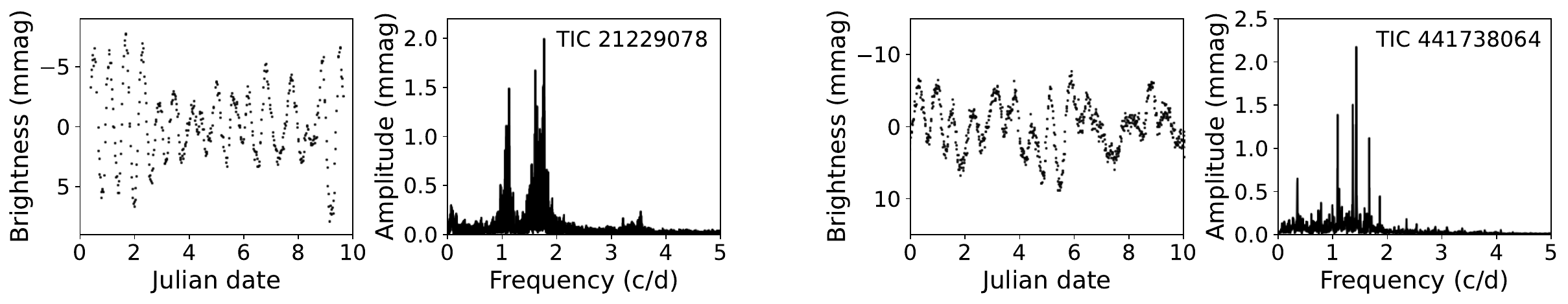}} \\ \hline

{\Large GDOR+DSCT; DSCT+GDOR} & 
(ir)regular variations, beating, bumps, interference, sharp variations &  
two or more independent peaks below and above 5\,c/d & 
simultaneous p- and g-mode pulsations \\
\multicolumn{4}{c}{\includegraphics[width=1.0\textwidth]{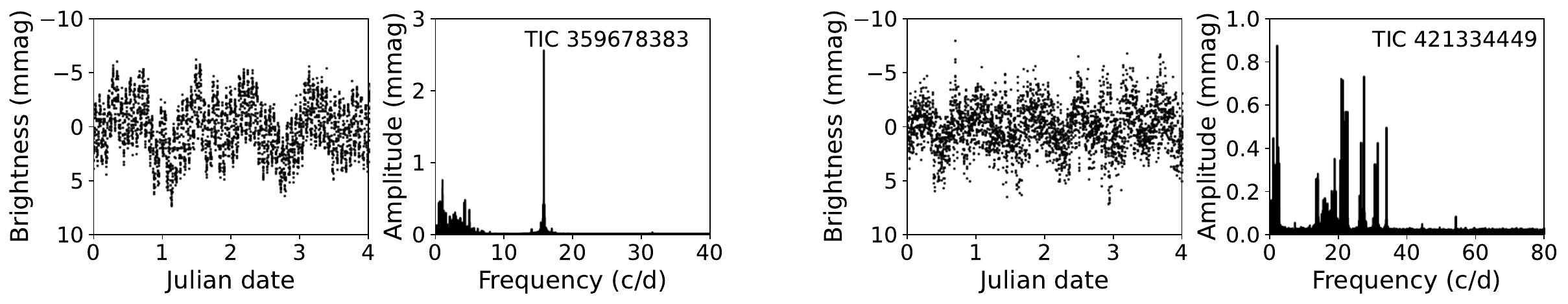}} \\ \hline
\hline                  
\end{tabular}
\end{adjustbox}
\end{table*}

%
\renewcommand{\arraystretch}{1.2}
\setlength{\tabcolsep}{4pt}
\begin{table*}[htbp!]
\caption{Typical examples of the variability types caused by the rotation of a single star. The columns are the same as in Table~\ref{Tab:VarTypes_Bins}. For the type 'ROT' we show the phase curves because the variations are more visible.}             
\label{Tab:VarTypes_ROT}      
\centering  
\begin{adjustbox}{width=\textwidth,center}
\begin{tabular}{>{\centering}m{0.22\textwidth}CCC}
 
\hline\hline       
                     
Type & 
{\normalsize  Light curve} & 
{\normalsize Frequency spectrum} & 
\normalsize Physical origin \\ 
\hline      

{\Large ROTM} & 
strictly repeating pattern, smooth variation without sharp features, maxima and minima generally different, superposition of two waves & 
one or two dominant peaks that are harmonics of the basic rotational frequency ($f_{\rm 2}=2f_{\rm 1}$), low-amplitude harmonics of $f_{\rm 1}$ may be present & 
rotation of a star with abundance anomaly spots \\ 
\multicolumn{4}{c}{\includegraphics[width=1.0\textwidth]{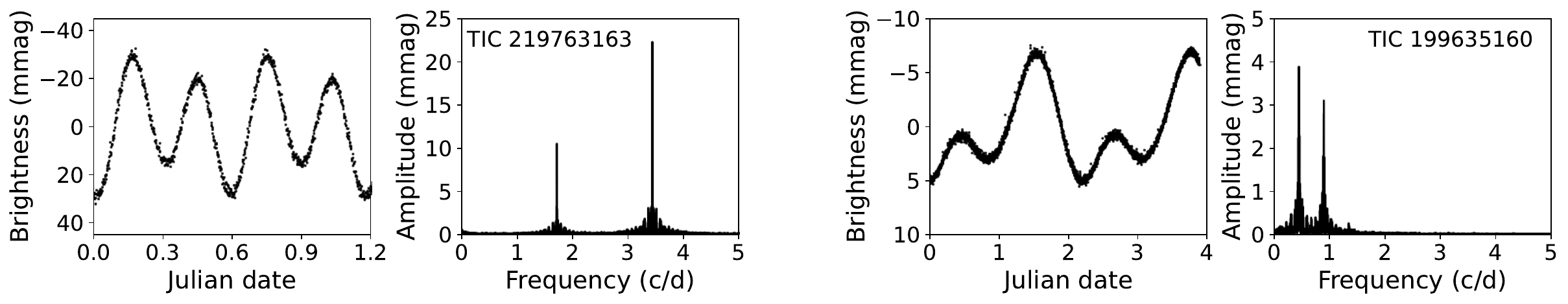}} \\ \hline

{\Large ROTS} & 
semi-regular variations superimposed on a basic periodic pattern & 
groups of (unresolved) peaks at positions close to harmonics of the strongest peak & 
rotation of a star with migrating (and forming or disappearing) spots, activity similar to our Sun, possible instrumental or data reduction artefacts \\
\multicolumn{4}{c}{\includegraphics[width=1.0\textwidth]{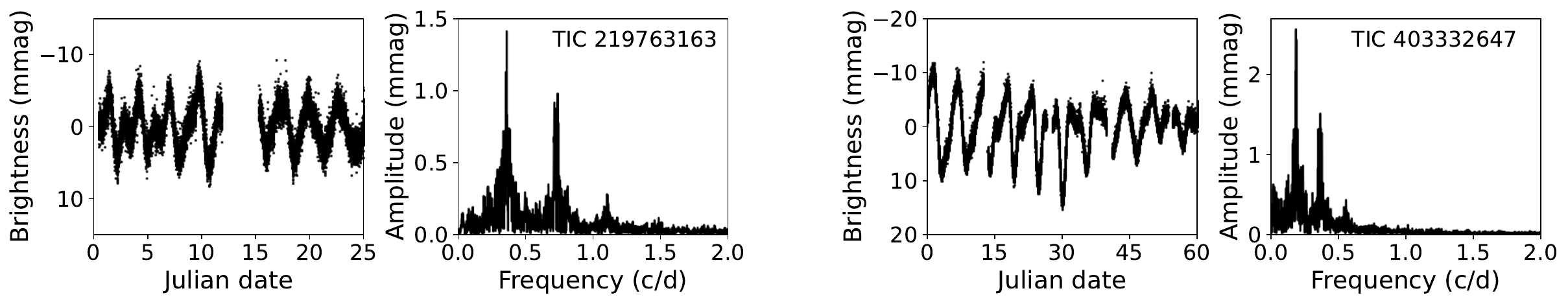}} \\ \hline

{\Large ROT} & 
repeating stable features & 
harmonics of the strongest frequency & 
likely some phenomena related to the rotation of the star \\
\multicolumn{4}{c}{\includegraphics[width=1.0\textwidth]{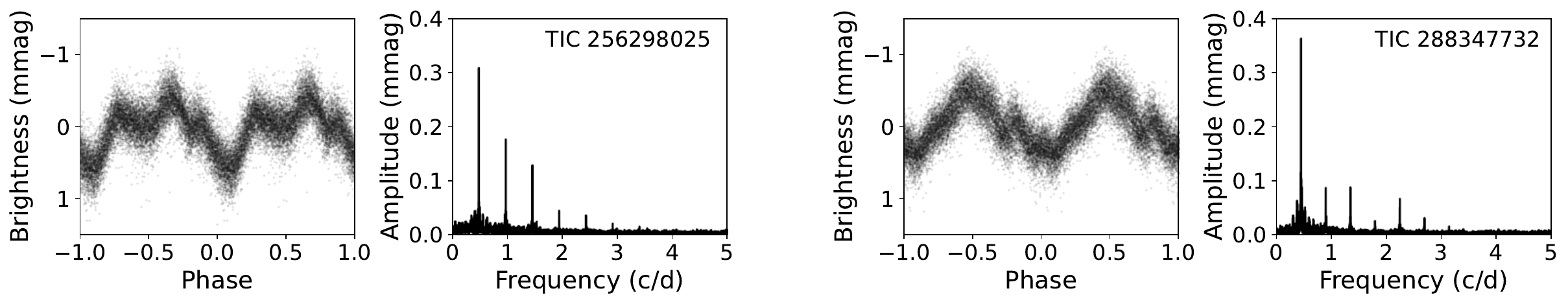}} \\ \hline

\end{tabular}
\end{adjustbox}
\end{table*}

%
\renewcommand{\arraystretch}{1.2}
\newcolumntype{M}{>{\centering\arraybackslash\small}c}
\setlength{\tabcolsep}{6pt}
\begin{table*}[htbp!]
\caption{Sample of our classification table.\ Shown are the designation in the TIC catalogue (column 1), positions (columns 'RA' and 'Dec.'), classification of the stars ('TYPE'), spectral type based on the LAMOST spectra, dominant frequency ('$f$'), zero epoch '$M_{0}$' (for eclipsing binaries), amplitude of the brightness variation ('$\Delta T$'), and the designation and classification in the VSX catalogue (VSX and VSX type) of the sample stars. The column 'Blend' gives a TIC number of the blending star that is variable, while the last column '$N_{Stars}$' gives the number of stars that are closer than 5 pixels from the object and the difference between object and the blending star is less than 5\,mag. The full is only available in electronic form
at the CDS.}             
\label{Tab:Main}      
\centering  
\begin{tabular}{MMMMMMMMMMMM}
 
\hline\hline       
                     
TIC & RA (deg) & Dec. (deg) & TYPE & Sp. type & $f$ (c/d) & $M_{0}$ & $\Delta T$ (mmag) & VSX & VSX type & Blend & $N_{stars}$ \\ \hline      

21002602 & 272.7745522 & 53.4938379  & ROTS & & 0.376 & & 4.0 & & & & 0\\
21018571 & 273.0771419 & 52.66231947 & ELL  & &0.993 & & 2.0 & & & & 0\\
21031802 & 273.3209365 & 52.41759483 & DSCT & & 9.391 & & 10.0& & & & 3\\
... & ... & ... & ... & ... & ... & ... & ... & ... & ... & ... & ...\\
\hline

\end{tabular}
\end{table*}

\section{Ambiguity in classification and other related shortcomings}\label{Sect:AmbiguityClassification}

\subsection{Blending}\label{Subsect:Blends}

One of the reasons why we give only a rough estimate of the amplitude in Table~\ref{Tab:Main} is that the value can be significantly affected by stars contaminating the light in the apertures \citep[TESS has  a spatial resolution of only 21"/px,][]{Ricker2015}. In an extreme case, the star in the sample itself can be stable and the variation can come from a nearby variable star. We did not investigate this issue in detail but we performed a quick analysis concerning stars within our sample. We checked stars that were closer than 10 pixels from each other. Based on the amplitude of the peaks in the frequency spectrum, we decided which star is variable and which is not. 

\begin{figure}
\centering
\includegraphics[width=0.47\textwidth]{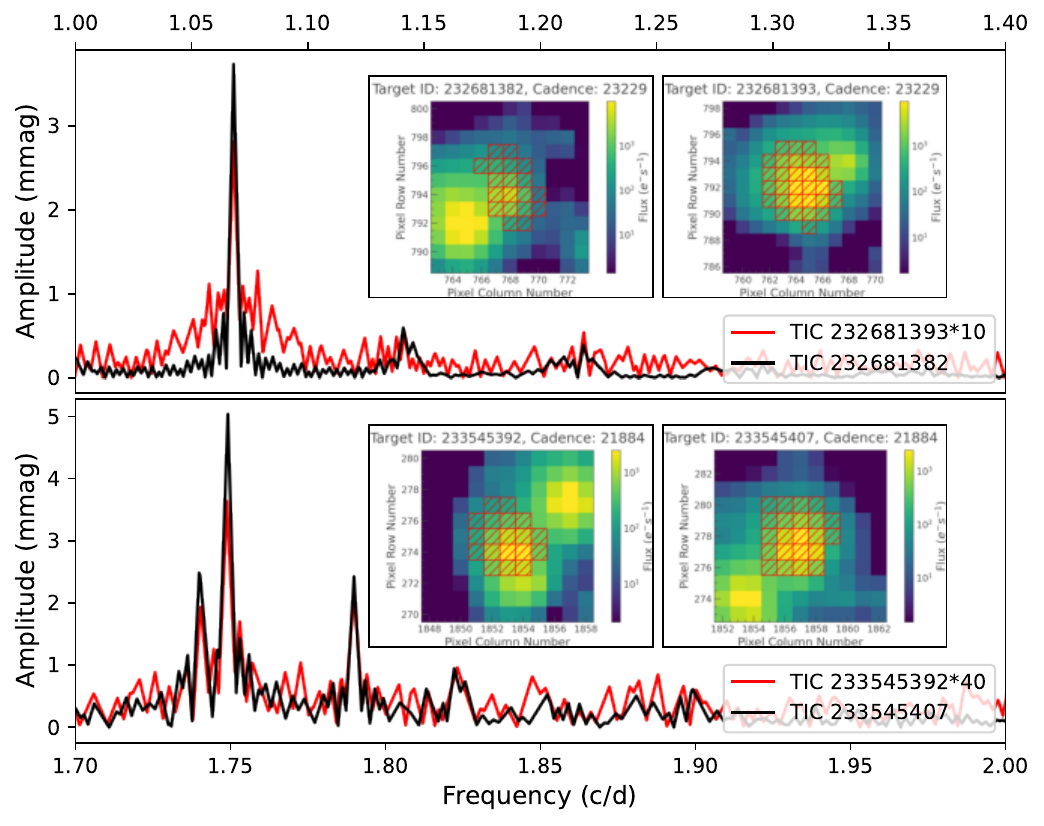}
\caption{Frequency spectra of two couples of blending stars. The apertures are shown in the insets. The red spectra correspond to non-variable stars contaminated by the nearby variables. The amplitudes of the peaks of non-variable blends are multiplied for a better visibility, by 10 and 40.}
\label{Fig:Blends}
\end{figure}

Figure \ref{Fig:Blends} shows two examples of blended stars. In the top panel, we can see that the variability comes from the fainter star, TIC~232681382, which has peaks in the FT with higher amplitudes. In the bottom panel, the situation is the opposite: the brighter star (TIC~233545407) is the variable. In our sample, we identified 29 such couples among stars classified as variable from which 21 pairs show the same FT with different amplitudes. For these cases, we comment in the last column of Table~\ref{Tab:Main} that the variability is caused by a close companion. 

There might be more blends with fainter stars, but a detailed inspection of the stars around each object in our sample is beyond the scope of this paper. Nevertheless, we cross-matched our sample with the GAIA DR2 catalogue \citep{GAIA2018} and identified all stars that are less than 5\,mag fainter (in GAIA $g$ filter) than our variable star and are closer than 105" (5\,px) to it. We give the number of such stars in the last column of Table~\ref{Tab:Main}. This value gives at least a warning about possible blends that can also cause false positive identifications of variable stars.

\subsection{Instrumental, data reduction artefacts, and semi-regular and long-period variations}\label{Subsect:InstrumentalEffects}

The data we used is the SAP processed with the PDC pipeline \citep{Smith2012,Stumpe2014}. This pipeline uses co-trending basis vectors to remove systematics such as those resulting from the focus changes, spacecraft pointing jitter, and other stochastic errors, as well as the crowding problems \citep[see e.g.][]{Jenkins2016,Jenkins2020,Kinemuchi2012}. Mostly because of scattered light and due to improper aperture definition, there might be additional light variations present in the data (as already pointed out in Sect.~\ref{Sect:Identification}) that is not intrinsic to the star. It is not always possible to distinguish between stellar variation and instrumental and reduction artefacts. 

Generally, we expect that the instrumental/reduction artefacts will cause irregular or semi-regular variations with periods longer than 1 day producing groups of low-amplitude peaks with frequencies below 1\,c/d in the FT. Typically, artefacts can be confused with stellar activity, rotational modulation due to spots, or GDOR type pulsations. From the high risk of instrumental artefacts or misclassification, we conservatively assigned spurious cases with a 'VAR' label although some of the stars could be assigned with a particular variability class. Actually, stars showing similar light curves and FTs as TIC~334678134 and TIC~417734885 (shown in the top row of Table~\ref{Tab:VarTypes_Puls}) have been often considered as stars showing rotational modulation or stellar activity \citep[e.g.][]{Balona2011rot,Balona2013,Bradley2015}.

\subsection{Pulsating variable stars}\label{Subsect:PulsVars}

There are three types of pulsators in the region of A-F stars near the main sequence. 
First of them are DSCT stars ($p-$mode pulsators) that are easily recognizable and the class is well defined because basically any variability with frequencies above 5\,c/d is caused by pulsations of DSCT type (for examples see the figures in Table~\ref{Tab:VarTypes_Puls}). There are also Population II stars that show similar behaviour as DSCT stars (SX Phe stars). However, we do not distinguish between these two types. 

Regarding the FT, the upper limit of 100\,c/d differentiates DSCTs from the second pulsating type,
roAp stars \citep{Kurtz1982}. However, the upper limit is not firmly set and depends on the particular author. The mechanism responsible for the rapid oscillations is not yet known. Currently the most accepted is the opacity mechanism in the hydrogen ionisation layer where the convection is suppressed by the magnetic field \citep{Balmforth2001}. Recently, \citet{Balona2022} found that high frequencies appear also in non-magnetic stars, while DSCT frequencies were found in magnetic stars where they should be damped \citep{Saio2005,Murphy2020}. \citet{Balona2022} also found that there is no frequency limit differentiating DSCT from roAp frequencies and argues that there is no need for the roAp class.

The last type of pulsations occurring among A-F type main sequence stars are pulsations of GDOR type. The g-mode pulsations are believed to be excited by the convective flux blocking mechanism \citep{Guzik2000,Dupret2005}. Frequencies observed in GDOR pulsators are usually between 0.3 and 3\,c/d \citep{Kaye1999,Henry2011}, but can also be above 5\,c/d, although with small amplitudes \citep{Grigahcene2010}. Similarly, the DSCT pulsations can have frequencies below 5\,c/d but with small amplitudes \citep{Grigahcene2010}. This can lead to a misclassification between GDOR and DSCT stars in a small fraction of stars. We classified a star as a GDOR if it shows two or more peaks with frequencies below 5\,c/d that are not harmonics of each other. Peaks in the FT of GDOR stars often appear in groups \citep{Balona2011gdor,Saio2018}. Such a pattern is characteristic for a significant portion of GDOR stars (see the examples in Table~\ref{Tab:VarTypes_Puls} and Fig.~\ref{Fig:GDORMosaic}). We cannot exclude the possibility that in clear GDOR stars some of the peaks can be caused by spots and rotation. 

A significant fraction of the A-F pulsators show both GDOR and DSCT pulsations (see the examples in Table~\ref{Tab:VarTypes_Puls} and in Fig.~\ref{Fig:HYBSosaic}). Some of them were known prior to space missions \citep[e.g.][]{Henry2005,Handler2009}. These stars are called hybrids. \citet{Balona2014} found that basically all DSCT stars show low-frequency peaks and are, therefore, hybrids, although some exceptions do exist \citep{Bowman2017}. We marked a star as a hybrid when at least two significant peaks ({\it S/N}$>4$) with frequencies below and above 5\,c/d are present. This is a somewhat weaker criterion than has been used, for example, by \citet{Uytterhoeven2011} and \citet{Bradley2015} who used an amplitude threshold and required that the amplitudes of the peaks in GDOR and DSCT regimes are similar (factor of 5-7). There is a risk that some of the frequencies can actually be a combination of other frequencies \citep{Balona2014,Bowman2018}. This could lead to a wrong hybrid classification. Further, in rapidly rotating GDOR stars, some of the peaks above 5\,c/d can actually be $g$ modes shifted to higher frequencies due to rotation \citep{Bouabid2013}. However, we assume that these issues would be present only in a small fraction of the stars. Because a detailed individual frequency analysis of the data is beyond the scope of this paper, we have not considered these effects in our classification.

A more serious problem for the classification and identification of the real highest peak in the FT is the presence of sub-Nyquist artefacts. In case of perfectly sampled data with a 30-minute cadence, the value of the Nyquist frequency would be $f_{\rm Nyquist}=24$\,c/d. Frequencies of some of the pulsation modes of DSCT stars can lie above this limit. A frequency $f_{\rm Real}$ in range of $24<f_{\rm Real}<48$\,c/d will be reflected to a position $f_{\rm Reflected}=2f_{\rm nyquist}-f_{\rm Real}=48-f_{\rm Real}$. DSCT frequencies close to $2f_{\rm Nyquist}$ will be reflected to a low-frequency region, where they possibly could mimic GDOR and ROTS frequencies. However, such frequencies will be heavily damped in amplitude, since the corresponding period of the pulsation cycle is comparable to the exposure time of the LC data \citep{Balona2014}. We performed a simple test and compared the FT of LC and SC data and found out that about half of the GDOR candidates show a peak with higher amplitude close to $2f_{\rm Nyquist}$ than close to 0\,c/d in the LC data. However, the SC data showed in all cases that the frequencies below 5\,c/d are the real ones and that there is no considerable risk that the GDOR stars are confused with DSCT stars due to Nyquist reflections. 

Concerning the DSCT stars, the SC data ensure that the DSCT frequencies are identified properly and the classification is correct. In the LC data, the perfect sampling is disrupted by the downlink of the data, which causes the observations to not restart after exactly $n\times$30\,minutes (where $n$ is an integer). This de-phasing is boosted in the TESS CVZ by multiple gaps (downlinks) and should allow for a reliable identification of the real pulsation frequencies of DSCT stars even if they are higher than $f_{\rm Nyquist}$ \citep[][]{Murphy2015}. However, the identification is not always perfect. 

\begin{figure*}
\centering
\includegraphics[width=0.98\textwidth]{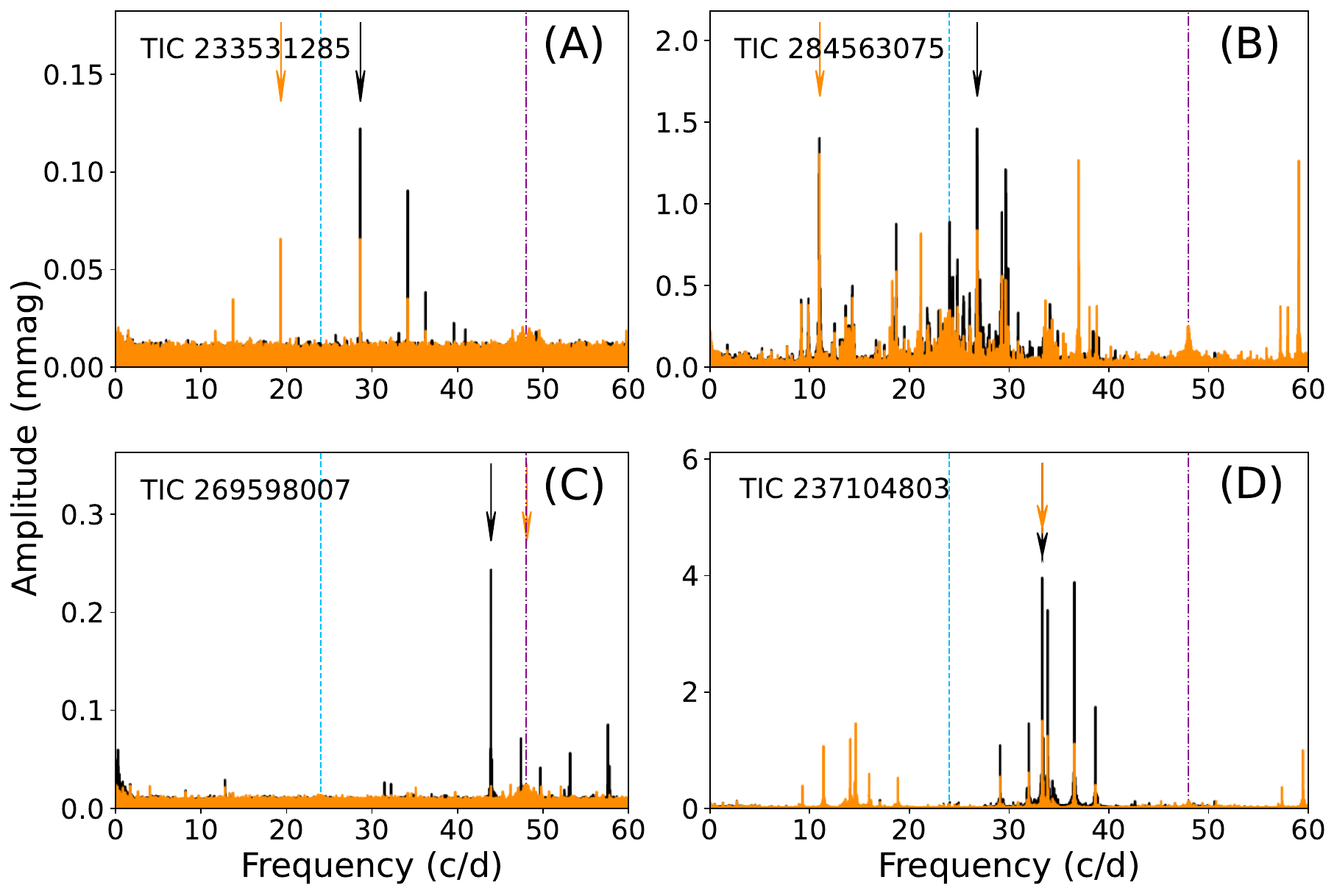}
\caption{Comparison of the FT based on the LC (orange) and SC (black) data for four DSCT stars. Frequencies with the highest amplitudes in the LC and SC data are marked with arrows of corresponding colour. The Nyquist frequency and its harmonic at $2f_{\rm Nyquist}$ are shown with the vertical light blue and magenta lines, respectively. Panel (A) shows the example when the Nyquist reflection of the highest peak would be detected in LC data, panel (B) shows the situation when the peak with the second highest amplitude would be the highest in the LC data, and panel (C) illustrates a situation when almost no variation is seen in the LC data and the identification completely failed in LC data. Panel (D) shows an example of when the proper peak would also be identified in LC data.}
\label{Fig:Nyquist}
\end{figure*}

For stars classified as DSCT based on the LC data, we performed the FT analysis again in the range 0-60\,c/d to retrieve possible frequencies above 24\,c/d. To check how reliable these frequencies are in the LC data, we compared the results of 16\,\% of all DSCT that have both LC and SC data (examples are shown in Fig.~\ref{Fig:Nyquist}). The frequency with the highest amplitude identified in the SC data was successfully retrieved in only 50\,\% of the test cases in the LC data (see panel (D) in Fig.~\ref{Fig:Nyquist}). Thus, the DSCT frequencies above 24\,c/d in Table~\ref{Tab:Main} may be wrong in 50\,\% of stars that do not have SC data. However, the classification as DSCT remains correct. Furthermore, we noticed that the frequency with the highest amplitude in the FT of SC data (that we expect to be the real dominant frequency) might not necessarily be the dominant frequency in the FT of LC data (panels (A)-(C) in Fig.~\ref{Fig:Nyquist}). This is probably the result of the different sampling behaviour and smaller amplitude of the fast light variations in LC data. 

\subsection{Rotationally variable stars}\label{Subsect:RotVars}

Rotation is a stable phenomenon. Thus, in rotating stars, we expect regular pattern in form of harmonics ($kf_{0}$) of the basic rotation frequency ($f_{0}$). The rotational variability can be induced by orbital motion of tidally deformed components of a non-eclipsing stars (ELL type, the last row in Table~\ref{Tab:VarTypes_Bins}) or by spots on the surface of a single rotating star (Table~\ref{Tab:VarTypes_ROT}). Among A-F type stars, these spots are believed to form via gravitational settling of light elements (He) and radiative levitation of heavy and earth-rare elements (e.g. Si, Sr, Cr, and Eu) forming a class of chemically peculiar stars \citep[][]{Michaud1970,Preston1974}. If a strong dipole magnetic field (inclined with respect to the rotational axis) is present, the spots formed usually around the magnetic poles can produce a rotational variability \citep{Stibbs1950,Kochukhov2011}. 

The rotation periods of spotted stars are long -- mostly of the order of days \citep{Sikora2019} to hundreds of days or even years \citep{Mathys2017,Mathys2020}. With one-year-long time base (at best), we are confined maximally to tens-of-days-long periods. The light curves of magnetic chemically peculiar stars have mostly sinusoidal or doubly sinusoidal light curves \citep{Jagelka2019}. We assign stars with stable light curves showing two (or more) harmonics in the FT as 'ROTM' (meaning magnetic rotators, figures in the top row of Table~\ref{Tab:VarTypes_ROT}) since we assume that the variation is caused by stable chemical spots that require strong magnetic fields\footnote{\citet{Jagelka2019} show that the role of the magnetic field in forming the chemical spots may not be as important as is formerly supposed.}. Possible additional long-term variations are interpreted as instrumental/reduction artefacts (see Fig.~\ref{Fig:Comparison}). We do not assign sinusoidal variation (one significant peak in FT) as ROTM since it cannot be unambiguously distinguished from ELL type, and we classify these stars as ROTM|ELL (see Sect.~\ref{Subsect:ROTMvsELL}).

As shown by many authors \citep[e.g.][]{Balona2011rot,Balona2013,Hummerich2018,Sikora2019}, A-F stars can also show more complex light curves including amplitude variations. The variation can be highly irregular and can cause variations from cycle to cycle \citep[see figures in the second row of Table~\ref{Tab:VarTypes_ROT} and also figures in e.g.][]{Balona2013,DeMedeiros2013,Sikora2019,Trust2020}. These variations are usually attributed to rotation and activity of a star with spots that can form and disappear and migrate due to differential rotation \citep[solar-like activity; e.g.][]{Balona2011rot,Debosscher2011}. However, one should be aware that mixing artefacts with real variations (see Fig.~\ref{Fig:Comparison}) can mimic time-varying spots.

\citet{Sikora2020} studied projected rotational velocities of 44 A- and late B-type stars showing rotational modulation in their light curves and found out that more than 10\,\%, but likely fewer than 30\,\% of main sequence A-type stars show this type of variability. This is surprising, since stellar activity of this type is not expected in hot stars without large convective envelopes. We assign stars showing semi-regular variations producing groups of (un)resolved peaks at positions of harmonics in the FT as ROTS -- rotators of solar type (figures in the second row of Table~\ref{Tab:VarTypes_ROT} and Fig.~\ref{Fig:ROTSMosaic}). The groups of peaks are assumed to be the result of differential rotation.

Usually, authors assume that also semi-regular variations with small amplitude and variations producing one group of peaks (or a single peak) are signs of rotation or activity \citep[e.g.][]{Balona2011rot,Uytterhoeven2011,Bradley2015}. We are more conservative and define such stars as VAR (see also Section~\ref{Subsect:InstrumentalEffects}).

Because there are many stars that show regular periodic patterns in the light curve that transforms into harmonics in the FT, we define the last type of rotational variability that is labelled as ROT (the last row in Table~\ref{Tab:VarTypes_ROT}, Fig.~\ref{Fig:ROTSMosaic}). The nature of these variations is, however, unclear. The origin can be explained by stable spots or some co-rotating structures governed by high-order magnetic multipoles, as suggested by \citet{Mikulasek2020} and \cite{Krticka2022} for the hot, chemically peculiar stars. However, more likely, these stars are a subgroup of GDOR stars (see Sect.~\ref{Sect:Discussion}).

\subsection{Spotted stars or non-eclipsing binaries?}\label{Subsect:ROTMvsELL}

\begin{figure}
\centering
\includegraphics[width=0.22\textwidth]{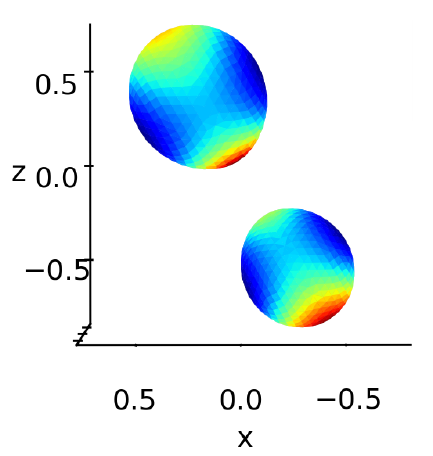}
\includegraphics[width=0.242\textwidth]{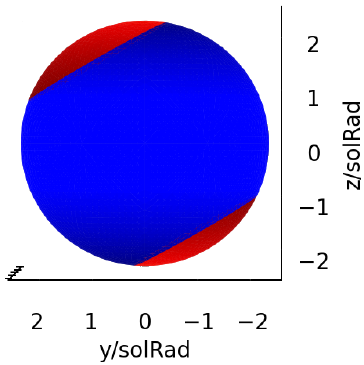}

\includegraphics[width=0.48\textwidth]{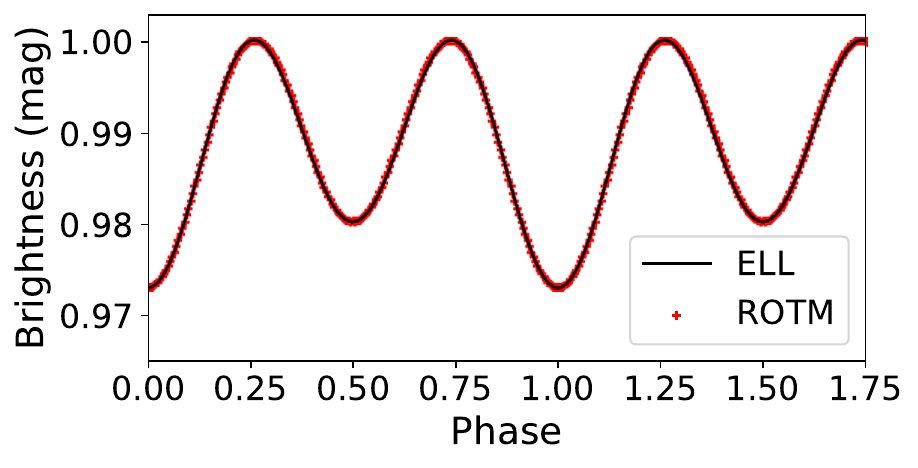}

\includegraphics[width=0.48\textwidth]{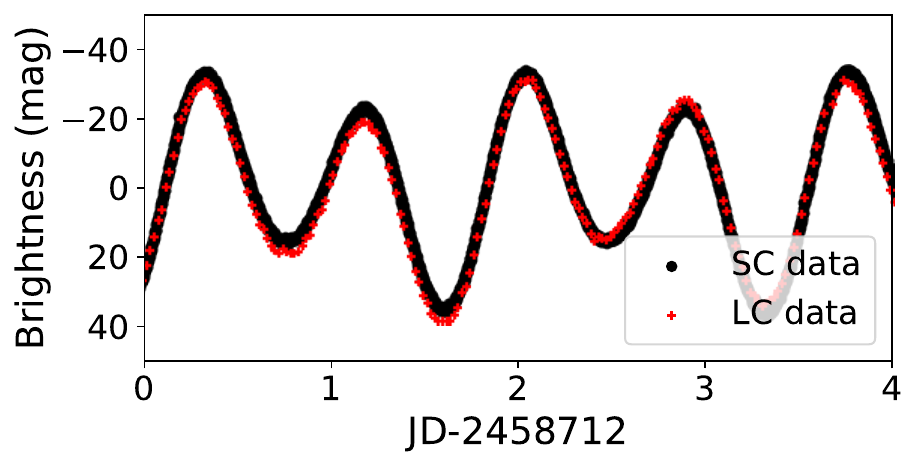}
\caption{Generated light curve (middle panel) assuming a non-eclipsing binary star with tidally deformed components (ELL, top left panel, black line in the middle panel) and a spotted star (ROTM, top right-hand panel, red crosses in the middle panel). An example of the light curve of a spectroscopically confirmed ELL star, IP Dra, that would rather be classified as ROTM is shown in the bottom panel. The models were generated using ELISa \citep{Cokina2021}.}
\label{Fig:Spotell}
\end{figure}

Without spectroscopic observations, based purely on single-passband photometry, there might be ambiguity between ROTM and ELL classes. In principle, the light variation of a spotted star (the top right-hand panel and red crosses in the middle panel of Fig.~\ref{Fig:Spotell}) can be equally described assuming gravity darkening caused by the tidal deformation of the components of a non-eclipsing system (the top left-hand panel and continuous line in the middle panel of Fig.~\ref{Fig:Spotell}). In addition, the amplitudes of the variations can be similar, as well as the periods and the FT. 

Due to geometric reasons, ELL stars must reach the same brightness in maximum light twice a period. However, the same variation can be reproduced assuming two spots with different sizes on the opposite sides of a star (middle panel of Fig.~\ref{Fig:Spotell}). Thus, this is not a decisive criterion for ELLs. On the other hand, different maximum light typical to ROTM stars can be easily modelled assuming one spot on one of the components of a non-eclipsing binary system. Thus, again, different height of maxima should not be taken as the decisive criterion between ROTM and ELL types but only as a hint. In addition, there can be artefacts that cause a deformation in the light curve, making the correct classification more difficult (Fig.~\ref{Fig:Comparison}). 

The ambiguity between ELL and ROTM can be demonstrated with a known ELL star IP Dra (the bottom panel of Fig.~\ref{Fig:Spotell}) that was spectroscopically confirmed to be a binary system \citep{Kjurkchieva2014}. The case of IP Dra shows the need of spectroscopic observations for reliable classification.

It was shown by \citet{Morris1985} and \citet{Beech1985} that if the light curve of ELL variables is described by the sum of sine and cosine functions ($m=A_{0}+\sum_{i=1}^{N} A_{i}\cos(i \Theta) + \sum_{i=1}^{N} B_{i}\sin(i \Theta)$, $\Theta$ being phase, $A_{i}$ and $B_{i}$ amplitudes, $A_{0}$ zero term), the term with the highest amplitude is $\cos (2\Theta)$. This criterion is sometimes used as a support for the assumption of the ellipsoidal variability \citep[e.g.][]{Dal2013,Li2021}. The light curve shown in Fig.~\ref{Fig:Spotell} also shows dominant $\cos (2\Theta)$. Since the dependence can be explained also with spots, the criterion with dominant $\cos (2\Theta)$ is not sufficient for classifying the star as ELL.

Stable chemical spots and rotational modulation are usually observed in hot stars and only exceptionally occur among stars with $T_{\rm eff}<7000$\,K \citep{Renson2009,Hummerich2018,Hummerich2021}. In addition, the amplitude of the brightness variations usually decreases towards longer wavelengths \citep[e.g.][]{Krticka2007,Krticka2015,Prvak2015}. Therefore, we expect only negligible incidence rate of ROTM type among stars with temperatures below 7000\,K and assign variations similar to those shown in Table~\ref{Tab:VarTypes_Bins} and Fig.~\ref{Fig:Spotell} as ELL. Stars hotter than 7000\,K showing rotational modulation with constant maximum brightness we assign as ROTM|ELL because of the ambiguity between these two types. 

\citet{Faigler2012} found seven binary stars that show light curves typical to ROTM stars and confirmed them spectroscopically. They explain the shape of the light curves as a combination of reflection effects, ellipticity and Doppler boosting \citep{Faigler2011}. Thus, it can easily happen that some of the stars classified as ROTM can actually be ELL. This example shows the unavoidable necessity of the spectroscopic observations in all ELL and ROTM stars to be properly classified.

\subsection{Pulsations or rotation?}\label{Subsect:PulsVsRots}

To our best knowledge, there are no definitive criteria how to distinguish between rotation and pulsations based purely on photometric data. The issue with discriminating between pulsations and rotation was already pointed out by many studies, such as \citet{DeMedeiros2013}, who claimed that a better selection of rotating variables can only be made using spectroscopic observations. \citet{Balona2011gdor} pointed out that irregular light curves can be attributed to slowly rotating stars and that the frequencies of the GDOR stars showing symmetric light curves or obvious beating are comparable to rotation frequencies. Actually, most of the studies dealing with the stellar classification based on the space data warn about the possible misclassification between pulsating stars and stars showing rotational variability \citep[e.g.][]{Uytterhoeven2011,Balona2011rot}.

The typical rotational frequencies calculated from the median projected rotational velocity $v\sin i$ and median radius $R$ for A-F stars are between 0.5 and 0.75\,c/d \citep{Royer2007}. However, the rotation frequencies can be as high as the critical rotational velocity, which corresponds to about $3-3.7$\,c/d in the fastest main sequence A-type stars \citep{Sikora2019}. Thus, there is a strong overlap of rotational variability and pulsations in the FT. 

\begin{figure*}
\centering
\includegraphics[width=0.98\textwidth]{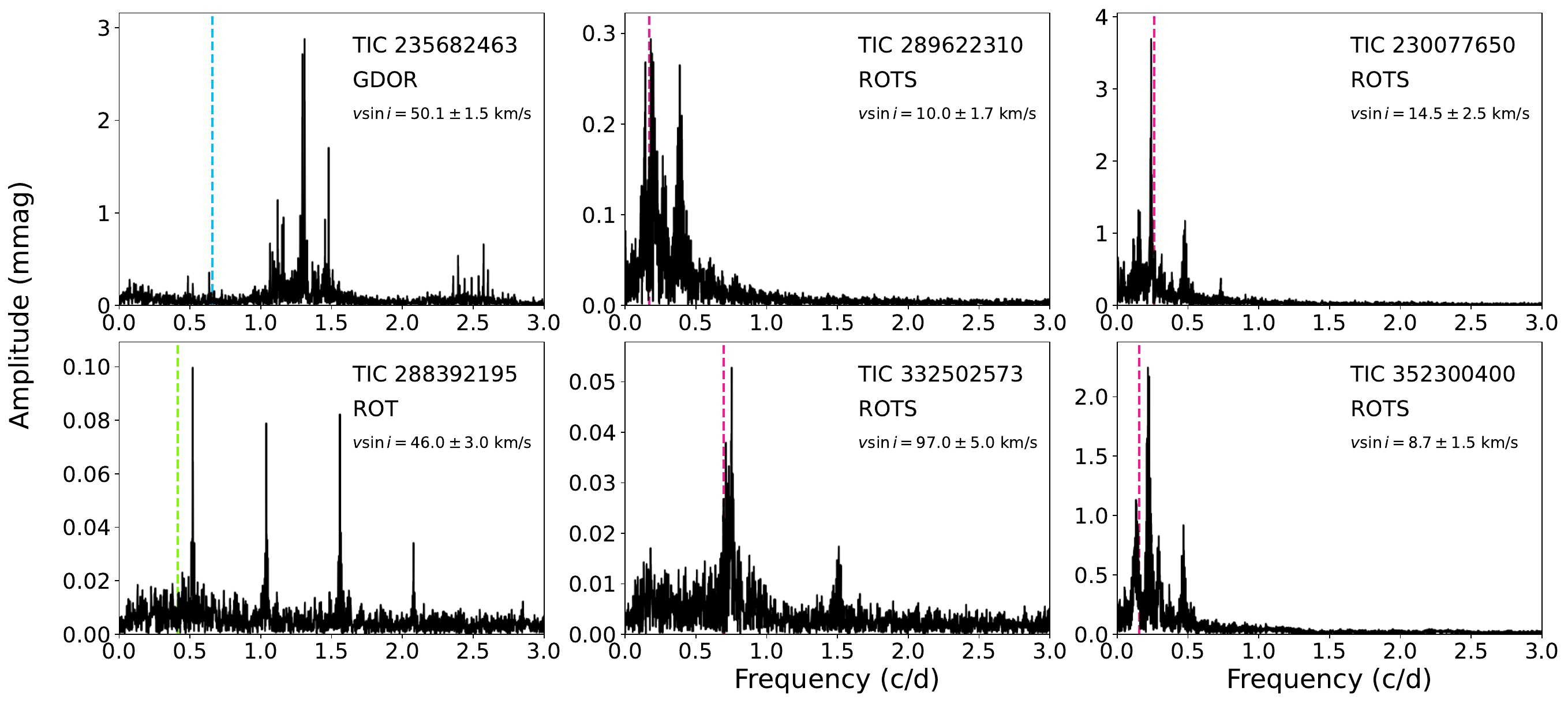}
\caption{Examples of FTs of GDOR (top left panel), ROT (bottom left panel), and ROTS stars (middle and right panels), showing also the position of frequencies calculated from observed $v\sin i$ and catalogue radii of the stars (vertical colour lines). It is apparent that the calculated frequency corresponds well with the observed peaks in the FT of ROTS stars.}
\label{Fig:Vsini}
\end{figure*}

We performed a very simple test inspired by \citet{Sikora2020}. We selected six bright examples of ROT, ROTS and GDOR stars and gathered spectra with the OES echelle spectrograph mounted at the Perek 2m telescope in Ond\v{r}ejov, Czech Republic (see Sect.~\ref{Subsect:Spectroscopy}). We modelled the spectra using {\textsc iSpec} and measured the $v\sin i$ (see Sect.~\ref{Subsect:Spectroscopy}). We calculated the bottom limit of the rotational frequency ($v\sin i \leq v_{\rm equatorial}$) by employing radii of the stars from the TIC catalogue \citep{Stassun2019}. The results are shown in Fig.~\ref{Fig:Vsini}.

Apparently, the measured frequencies agree well with the frequencies detected in the FT of the photometric data of the ROTS stars, while for GDOR and ROT stars the frequencies do not match well. However, the calculated frequencies shown with vertical dashed lines in Fig.~\ref{Fig:Vsini} are only lower limits. It is possible that some of the peaks shown in the left panels of Fig.~\ref{Fig:Vsini} can be addressed with inclined rotation axis and rotation. Thus, this test alone is not a proof that the variation in stars producing (un)resolved peaks at their harmonic positions in the FT are caused by rotation rather than pulsations. It is only strong support for this explanation. In addition, rotation of a star can shift the pulsation frequencies. An example how difficult the correct classification of such stars can be was published by \citet{Uytterhoeven2011b}. They found that variations in HD 171834, which produce unresolved peaks at the harmonic positions in the FT, are caused by rotation rather than pulsations.

\begin{figure}
\centering
\includegraphics[width=0.48\textwidth]{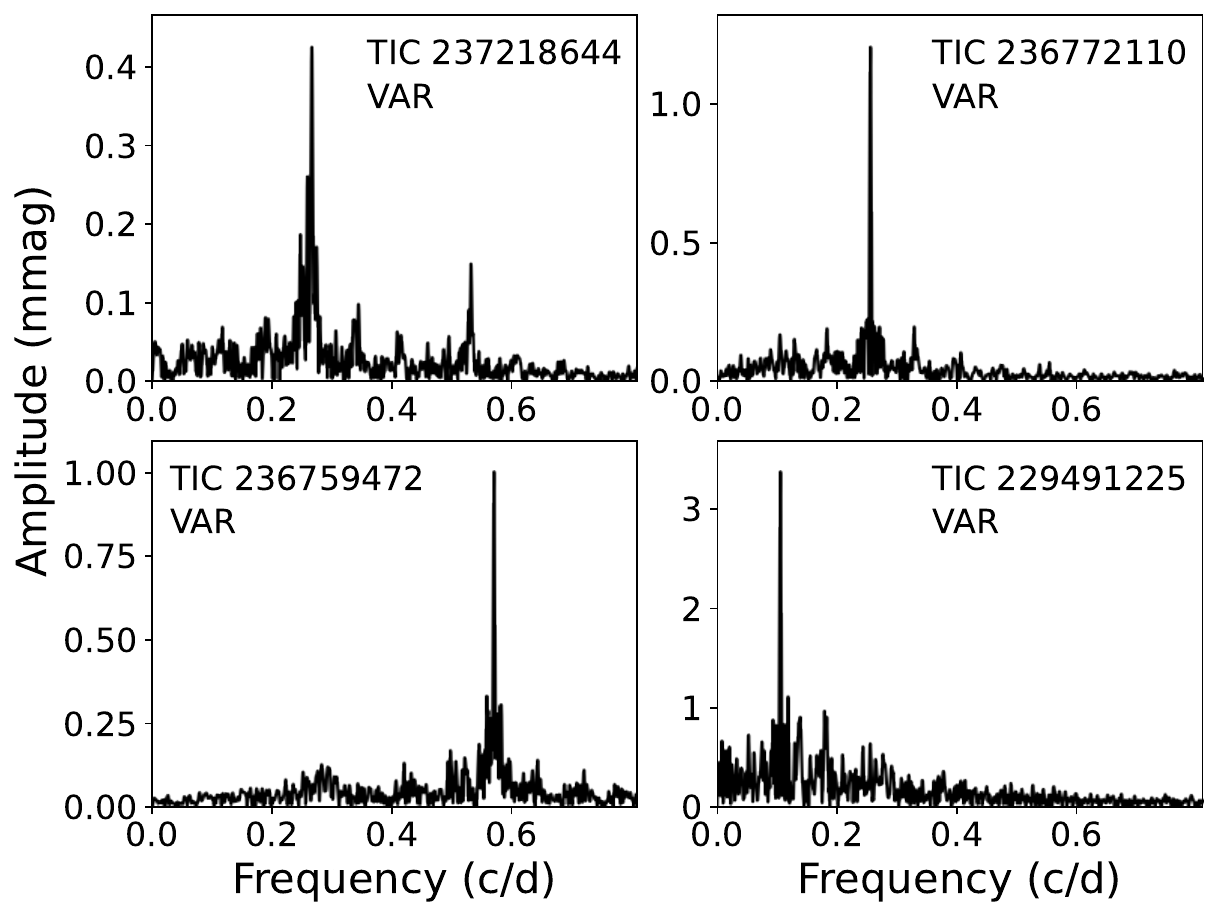}
\caption{Examples of frequency spectra of stars with ambiguous classification and, therefore, conservatively classified as VAR. See the text for details.}
\label{Fig:rotpuls}
\end{figure}

It is likely that there remain spurious cases in our sample that would need detailed investigation including spectroscopic observations. TIC~237218644 (top left panel of Fig.~\ref{Fig:rotpuls}) would be classified as ROTS according to our methodology, but the temperature of $9087\pm204$\,K \citep{Stassun2019} suggests that the origin of the variations is probably pulsations rather than rotation, since at such high temperatures, stellar activity linked with convection is not expected. There might also be uncertainty in temperature. Due to all these issues, we classify TIC~237218644 as VAR. The remaining panels of Fig.~\ref{Fig:rotpuls} contain stars that show only one significant peak in the FT. These stars could possibly be classified as ROTS but also as GDOR, ELL or ROTM types. We conservatively classify such stars as VAR. 

\section{Results and discussion}\label{Sect:Discussion}

\begin{figure}
\centering
\includegraphics[width=0.48\textwidth]{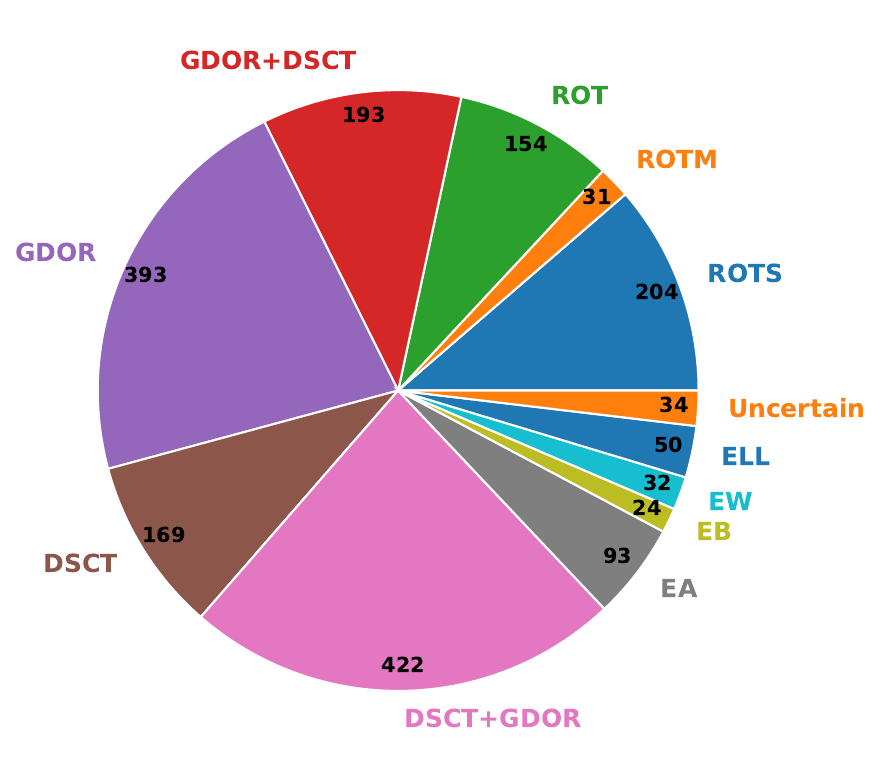}
\caption{Numbers of  stars in particular variability classes.}
\label{Fig:Types_stats}
\end{figure}

We identified 3025 stars (out of 5923 in our sample, 51\,\%) that show brightness variations. This is 12\,\% less than reported on the basis of preliminary results \citep{Skarka2021}. From the 3025 stars, we were able to assign 1813 stars (60\,\%) to a specific variability type. The number of stars in particular classes are shown in Fig.~\ref{Fig:Types_stats}. The most numerous are GDOR and DSCT classes and their hybrids (together 64.9\,\% of the variable stars). This is in excellent agreement with \citet{Uytterhoeven2011} who found 63\,\% of stars to show oscillations. Regarding eclipsing binaries (EA, EP, EA|EP, EB and EW classes), we found only 56\,\% of our binary candidates in the TESS eclipsing binary catalogue\footnote{\url{http://tessebs.villanova.edu/}} \citep{Prsa2022}. This discrepancy, however, comes from the fact that they used only 2-minute cadence. If we use only the SC data, then the agreement of our identification with the TESS EB catalogue is 95\,\%. 

In the 'Uncertain' class, there are 34 stars with ambiguous classification with two equally probable types, for example ELL|ROTM and EA|EP. It was not possible to reliably classify the rest of the variable stars, marked as VAR (1212 stars; see Sect.~\ref{Sect:AmbiguityClassification}). These are not shown in Fig.~\ref{Fig:Types_stats}. In addition to the numbers given in Fig.~\ref{Fig:Types_stats}, we identified 6 RR Lyrae stars (1 RRAB, 2 RRAB/BL, 3 RRC) and 7 heartbeat (HB) stars. In five stars we detected frequencies in the roAp range. However, it is not clear whether these are not combination frequencies. Not counting GDOR and DSCT hybrids, 50 stars show a combination of variability types, for example, EA+DSCT and GDOR+HB. 

\begin{figure*}
\centering
\includegraphics[width=0.48\textwidth]{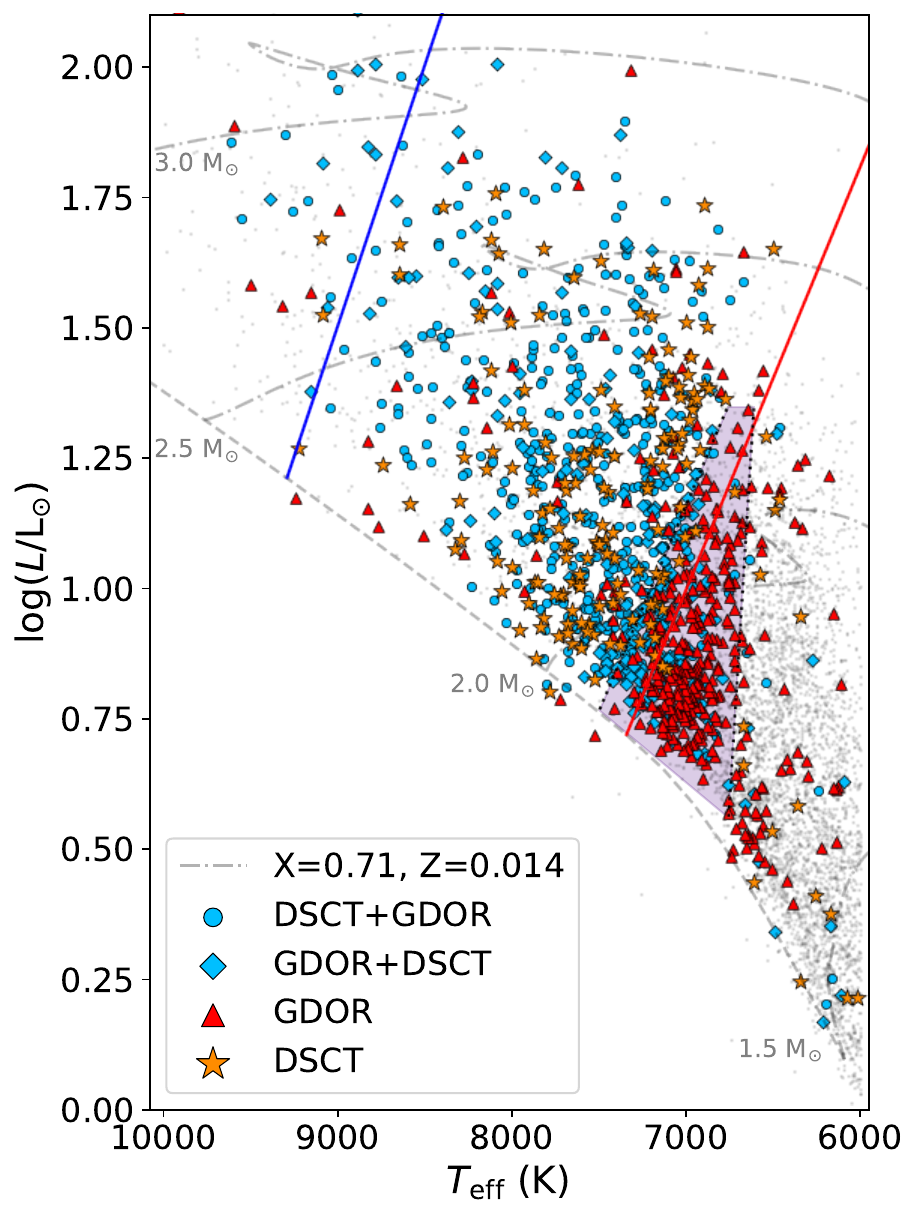}
\includegraphics[width=0.48\textwidth]{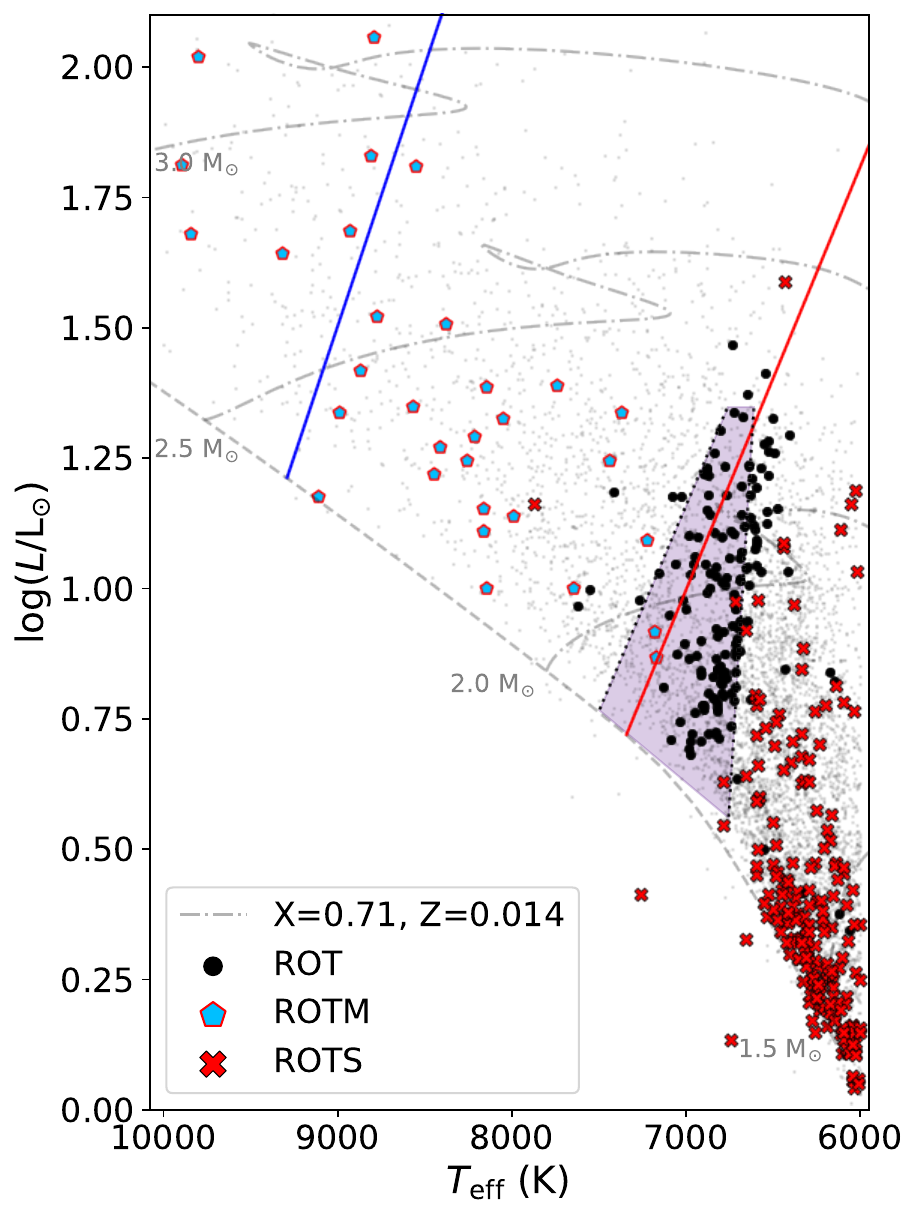}
\caption{Hertzsprung-Russell diagram showing pulsating stars (left panel) and rotationally modulated stars (right-hand panel). The dashed line shows the zero-age main sequence, dash-dotted lines show the evolutionary tracks for stars with different masses, all taken from \citet{Murphy2019}. The blue and red continuous lines show the empirical boundaries of the instability strip determined by \citet{Murphy2019}. The shaded area enclosed by the dotted lines shows the GDOR instability region, following \citet{Dupret2005}. The grey dots show all stars including non-variable stars in our sample.
}
\label{Fig:HRD}
\end{figure*}

\begin{figure}
\centering
\includegraphics[width=0.48\textwidth]{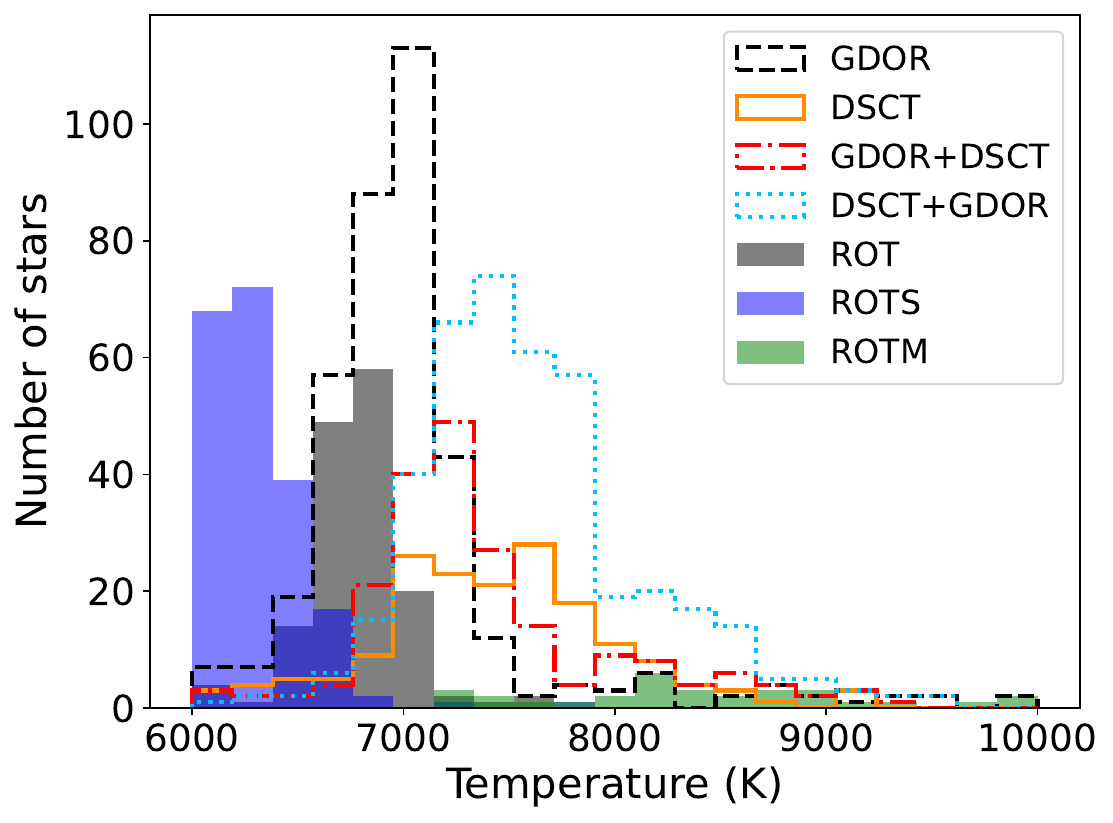}
\caption{Distribution of the stars with respect to their temperature.
}
\label{Fig:Hist_temp}
\end{figure}

Plotting all the stars in the Hertzsprung-Russell diagram\footnote{Temperatures are taken from \citet{Stassun2019}, luminosity of the stars calculated from temperatures and radii from \citet{Stassun2019}} (HRD; Fig.~\ref{Fig:HRD}) show some (expected) clumping. The vast majority of GDOR stars is located in a narrow part of the HRD well within the theoretical instability strip (IS) by \citep{Dupret2005}, as shown in the left panel of Fig.~\ref{Fig:HRD}. The distribution of GDOR stars has maximum at around 7000\,K and mean luminosity at about $\log L/L_{\odot}=0.8$ (Fig.~\ref{Fig:Hist_temp}). GDOR stars located outside the IS above approximately 7500\,K are uniformly spread over the whole temperature range. Different behaviour of stars in and out of the IS suggests that the hot stars marked as GDOR outside of IS are not of the same type although the nature of the variability is likely to be $g$-mode pulsations. 

The temperature distribution of DSCT and DSCT+GDOR hybrids is similar, although DSCT+GDOR hybrids are about 200\,K hotter on average. GDOR+DSCT hybrids have preferentially lower temperatures than their DSCT-dominant counterparts but have the same average temperature as DSCT stars (see Fig.~\ref{Fig:Hist_temp}). There are a few GDOR, DSCT, and hybrid stars with unexpectedly low temperature. We checked them and would all of them classify as they are (see Fig.~\ref{Fig:Cool}). The unexpected behaviour at low temperatures might indicate contamination of the light by nearby stars or less reliable temperatures, although the error bars are similar to other stars in our sample.  The groups of the rotating variables are well separated in the HRD. In addition, they have different temperature distributions. The ROTS stars (11.3\,\%) dominate the low-temperature part of the HRD, which can be naturally expected as the variations are supposed to be caused by the spots of solar type. The ROTM stars (1.7\,\%) are well spread out over the whole temperature range above 7000\,K. 

ROT stars (8.5\,\%) are almost exclusively located near and within the cool edge of the GDOR instability strip. This suggests that the observed brightness variations are likely connected with pulsations rather than with rotation. Alternatively, rotation and pulsation can be present simultaneously and ROT stars could be a subclass of GDOR stars. The GDOR stars have an average temperature of 7043\,K and a frequency of 1.358\,c/d, while ROT stars have a mean temperature of 6777\,K and a mean frequency of 0.578\,c/d. The mean amplitude is significantly lower for ROT stars (1.5\,mmag) than for GDOR stars (9.8\,mmag).

Our classification is in a very good agreement with the VSX classification. There are, of course, small differences but the variability types we provide are more specific or the same as in the VSX (with only a few exceptions). We applied our classification scheme to three of the previous studies dealing with the classification of variable stars based on the space data \citep{Uytterhoeven2011,Balona2013,Bradley2015} to see the differences in the classification paying special attention to stars with frequencies below 5\,c/d where the ambiguity between pulsation and rotation can emerge (see Sect~\ref{Subsect:PulsVsRots}). We downloaded the {\it Kepler} data of the stars with {\textsc Lightkurve} and performed the analysis exactly in the same way as for the stars in our sample.

\begin{figure*}
\centering
\includegraphics[width=0.48\textwidth]{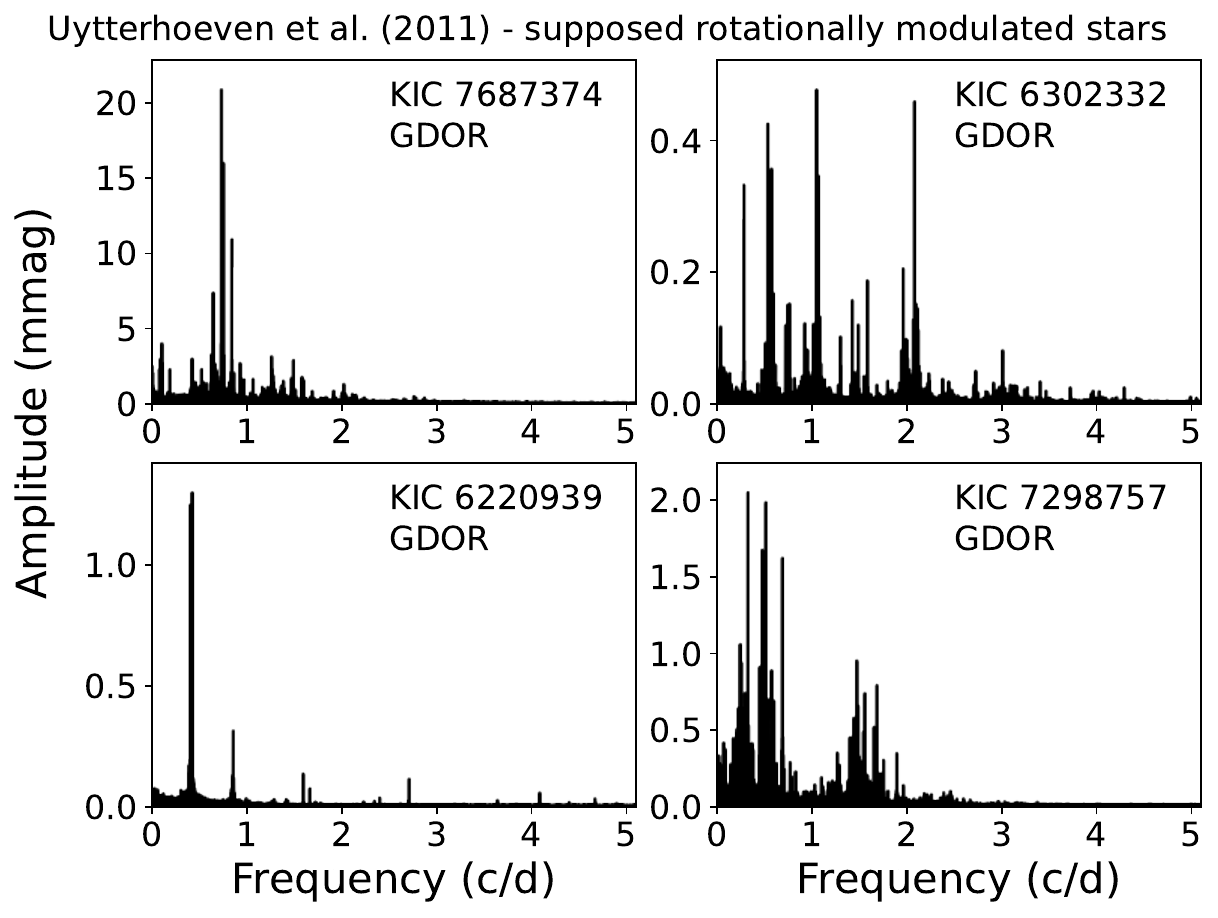}
\includegraphics[width=0.48\textwidth]{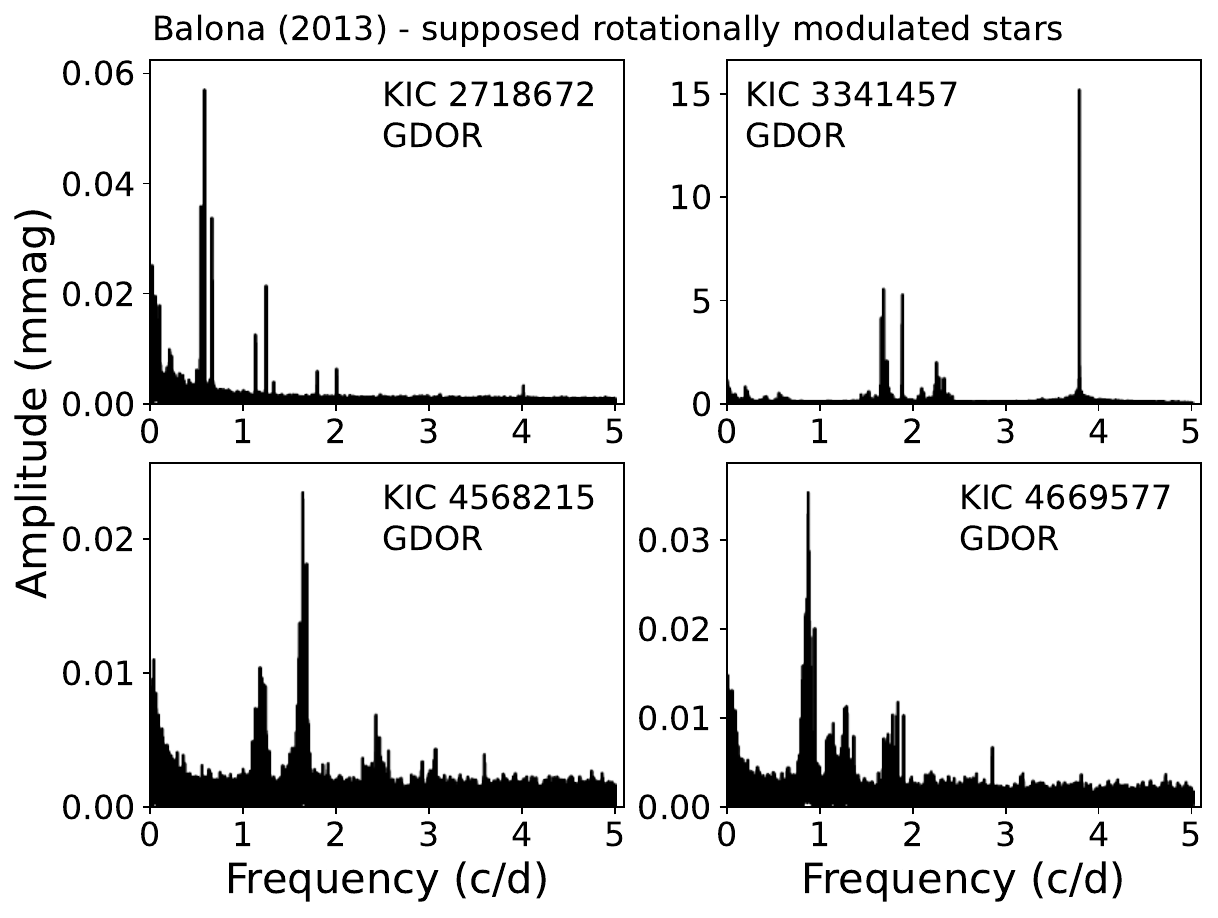}

\includegraphics[width=0.48\textwidth]{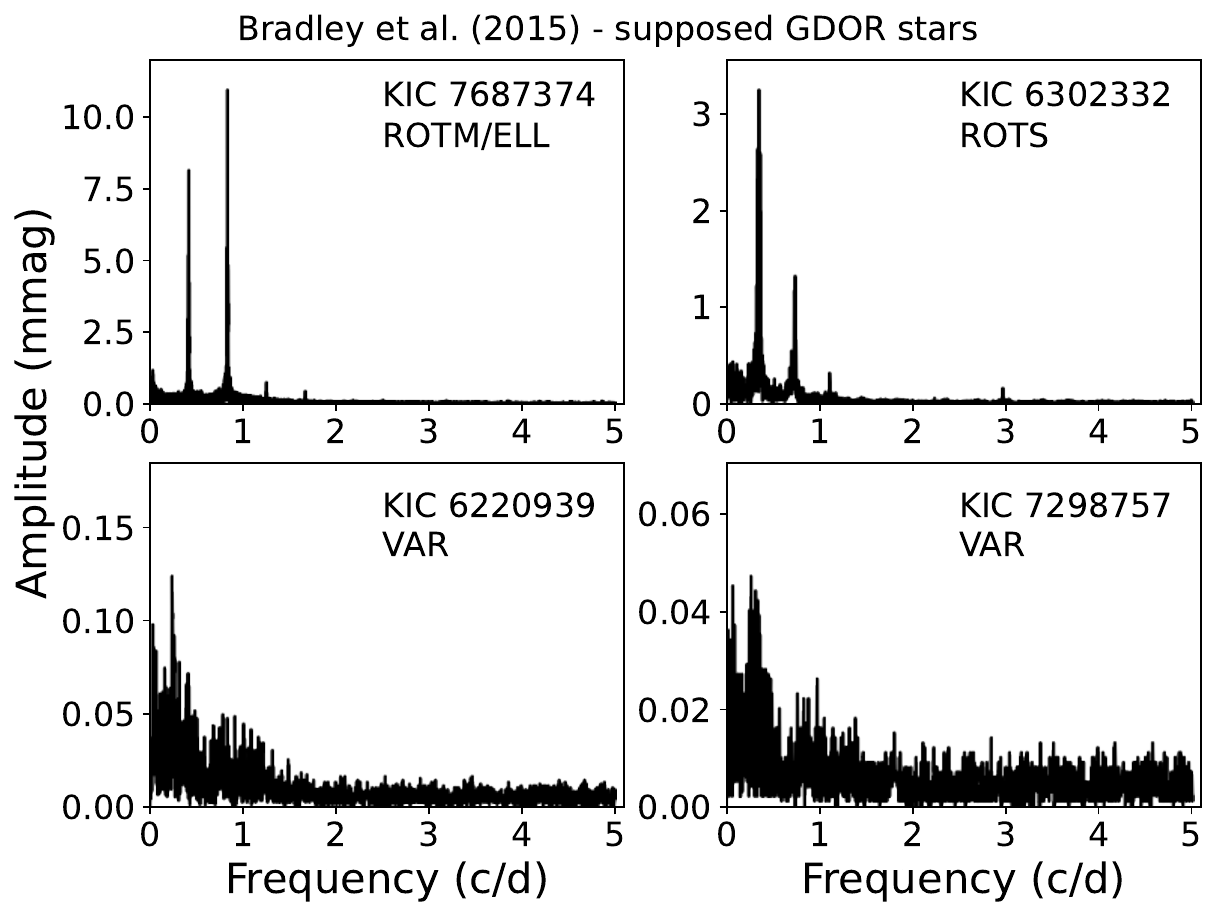}
\includegraphics[width=0.48\textwidth]{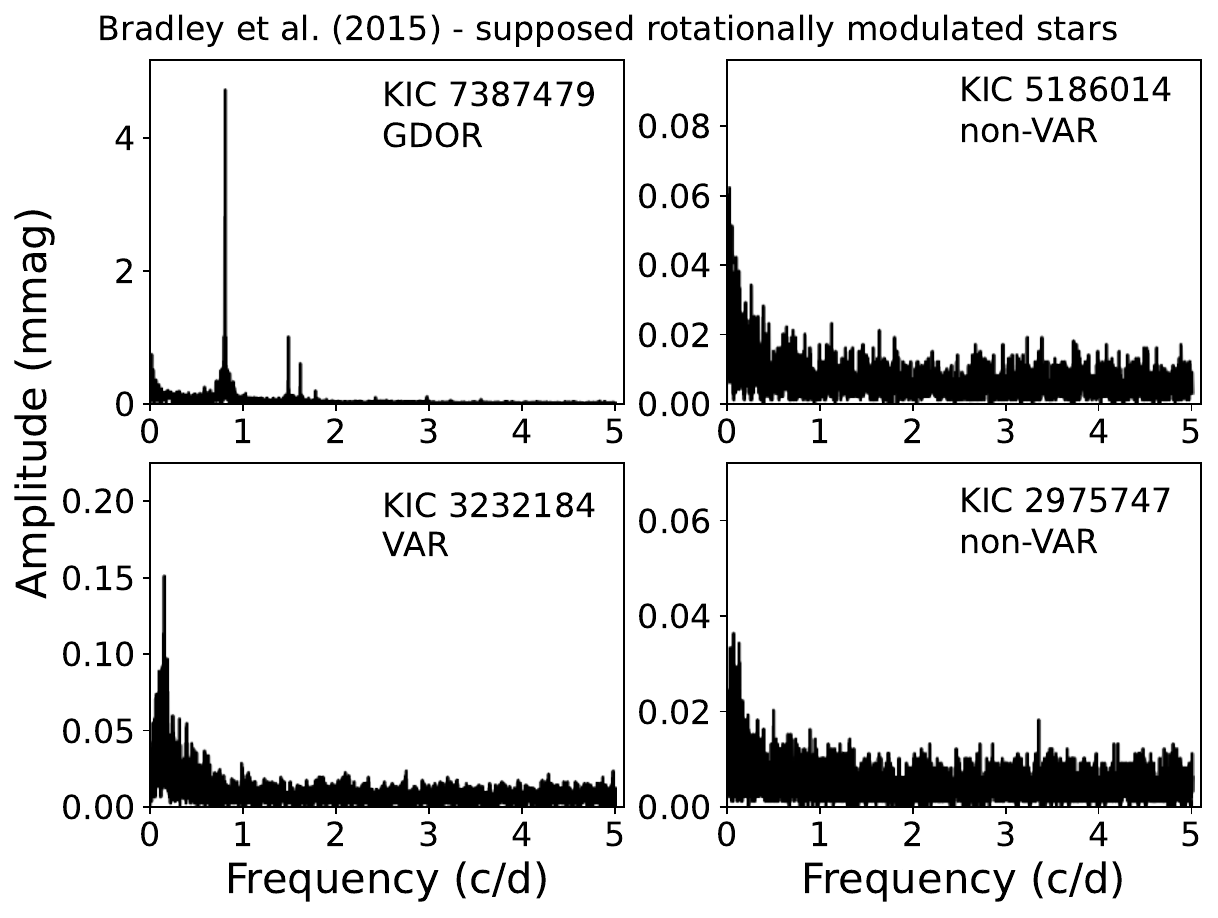}
\caption{Frequency spectra of stars observed by the {\it Kepler} mission and classified in previous studies (shown on the top of every four plots). The labels give how the stars would be classified using our methodology.}
\label{Fig:Discrepancy}
\end{figure*}

Among 87 stars classified by \citet{Uytterhoeven2011} as GDOR we found only 4 stars that we classified as ROTS and two stars that we classified as VAR. Thus, the agreement is excellent. However, in one third of stars classified by them as rotation or activity, we found that they are actually GDOR stars or hybrids (see examples in the four upper left-hand panels of Fig.~\ref{Fig:Discrepancy}). This is a significant discrepancy. In 25\,\% of the stars classified as rotational variables by \citet{Balona2013}\footnote{We checked only the first 200 stars from the list of \citet{Balona2013}.} we found GDOR variability (examples are shown in the four upper right-hand panels of Fig.~\ref{Fig:Discrepancy}) but we found perfect agreement for all 9 stars classified by \citet{Balona2013} as GDOR. 

The worst agreement with our classification is the classification by \citet{Bradley2015}. From their sample of 195 GDOR stars we classified 9 as non-variable stars, 39 as VAR, 29 as ROTS, 6 as ROT, 2 as ROTM/ELL and one clear RR Lyrae (four bottom left panels of Fig.~\ref{Fig:Discrepancy}). This means that we classified 44\,\% of stars marked by \citet{Bradley2015} as GDOR into a different variability class. We also checked 551 stars classified by \citet{Bradley2015} as rotating variables. We classified 403 of them (73\,\%) as VAR or non-variables although the variability of the majority of these stars can be caused by the stellar activity (see the examples in four bottom right panels of Fig.~\ref{Fig:Discrepancy}). The rest of the stars in their sample we classified as ROTM, ROTS and ROT stars with only insignificant contamination of GDOR. In general, variability of a significant part of stars in the VAR class can be actually caused by stellar activity, but these variations usually have small amplitudes and the classification is not trustworthy. If we assume that all the stars from the sample of \citet{Bradley2015} classified by us as VAR are rotational variables, then the agreement with their study would be very good.

Misclassification can be a serious issue in statistical studies and in training procedures based on machine learning and neural network methods. All the issues and discrepancies described above show that a commonly accepted methodology and classification system is desirable but difficult to be built. The border cases with ambiguous classification will always be present. It is the reason why our classification is rather conservative. Theoretical limits for the frequencies of GDOR pulsations, would be very helpful to distinguish between rotation and pulsations in a significant part of stars. It is worth to mention that we have not noticed any flare in the light curves of the sample stars.

\section{Conclusions and future prospects}\label{Sect:Conclusions}

We have performed a careful individual classification of 5923 A-F stars (temperatures between 6000 and 10000\,K) brighter than 11\,mag located close by and in the northern TESS CVZ. The classification is based on the TESS photometric data and the properties of the frequency spectra. We discussed the data, crowding, the effects of residual variations caused by improper data reduction, aperture definition, and the influence of these effects on the classification of the variable stars. We have also discussed differences and similarities between the classes and paid special attention to a proper classification. We did not aim to study variability classes in detail, nor did we highlight any particular star. We also did not show any of the interesting cases. These remain for future dedicated studies.

We adopted the VSX classification scheme. We did not deal with the morphological classification of the variable stars within the variability classes, as was done by previous authors \citep[e.g.][]{Balona2011gdor,Bradley2015}. Our classification of the intrinsic variability is based on the assumption of the basic physical phenomena, that is, rotation and pulsations. We introduced three classes of rotationally modulated stars: stars with stable chemical spots (ROTM), stars with regular brightness variations (ROT), and stars showing semi-regular variations assumed to be caused by the forming or ceasing of the spots in combination with differential rotation (ROTS).  

Our main results and findings can be summarised as follows:
\begin{itemize}
    \item The PDCSAP data with different cadences are generally different in both amplitude and cadence. Mainly the 30-minute data suffer from lots of residual variability that is not coming from the stars. This can lead to false positives and wrong classifications. 
    \item We provide a well-defined classification scheme and methodology for the proper classification of A-F variable stars based on the light-curve shape and the corresponding FT (Tables~\ref{Tab:VarTypes_Bins}-\ref{Tab:VarTypes_ROT}).
    \item We find that the identification of the real pulsation frequencies above the Nyquist limit based purely on the amplitudes of the corresponding peaks in the FT can be wrong in up to 50\,\% of cases. 
    \item We identify a new (sub)class of variables that show regular periodic light-curve shapes and harmonics of the basic frequency in the FT. First we assumed that these variations must be connected with rotation, but the position of these stars close to and within the GDOR instability strip suggests that the variations are likely caused by pulsations. ROT stars have longer periods (1.7 versus 0.74\,days), smaller amplitudes (1.5 versus 9.8\,mmag), and are cooler on average (6777 versus 7043\,K) than GDOR stars. The position of the stars in the HRD also suggests that the idea of co-rotating structures governed by high-order magnetic multipoles \citep[][]{Mikulasek2020,Krticka2022} likely cannot explain the variations in these relatively cool stars that are not expected to have strong stellar winds.
    \item Measurements of $v\sin i$ of four stars that show semi-regular brightness variations and unresolved groups of peaks near the harmonics of the basic frequency give a hint that the observed light variations may be linked with rotation rather than with pulsations. Stars in this group, called ROTS, can be mostly easily recognised from the GDOR stars.
    \item If the model of the light curve provides physically plausible results, there is no way to unambiguously distinguish between ellipsoidal variables and ROTM types purely on the basis of single-channel photometry. We have demonstrated this issue on IP Dra and generated models (Fig.~\ref{Fig:Spotell}).
\end{itemize}

On the comparison with previous studies, we have demonstrated that a commonly accepted classification scheme should be agreed upon. Our approach offers a well-defined system of classification with many examples. There still might be spurious cases, mainly regarding the GDOR, ROTS, and ROT classes; however, our sample should be clear of such ambiguous stars since we classified the spurious cases as VAR. The theoretical study of the limits of $g$-mode pulsations and all the possible patterns that can be expected in the FT would be very helpful in the future to better distinguish between rotation and pulsations. The only way to reliably classify the variable stars is via a detailed  case-by-case investigation that includes spectroscopic data. We warn about any blind usage of the (mainly LC) TESS data.

The approach presented in the paper is rather conservative. Our groups of stars provide good and reliable samples of stars without significant contamination suitable for statistical studies and the training of neural networks. We are planning spectroscopic observations and detailed investigations of particularly interesting stars identified during the analysis of the data. We will perform the same analysis for the southern TESS CVZ.

\begin{acknowledgements}
      MS, Pk, and RK acknowledge the support by Inter-transfer grant no LTT-20015. MM acknowledges the support by MEYS (Czech Republic) under the project MEYS LTT17006. This paper includes data collected with the TESS mission. Funding for the TESS mission is provided by the NASA Explorer Program. Funding for the TESS Asteroseismic Science Operations Centre is provided by the Danish National Research Foundation (Grant agreement no.: DNRF106), ESA PRODEX (PEA 4000119301) and Stellar Astrophysics Centre (SAC) at Aarhus University. We thank the TESS team and staff and TASC/TASOC for their support of the present work. We also thank the TASC WG4 team for their contribution to the selection of targets for 2-min observations. The TESS data were obtained from the MAST data archive at the Space Telescope Science Institute (STScI). We acknowledge the usage of the data taken with the Perek telescope at the Astronomical Institute of the Czech Academy of Sciences in Ond\v{r}ejov and would like to thank the observers for their work. We thank the referee for useful comments that improved the manuscript.
\end{acknowledgements}

\bibliographystyle{aa}
\bibliography{references}

\begin{appendix} 
\section{Examples of typical light curves and frequency spectra}

\begin{figure*}
\includegraphics[width=0.96\textwidth]{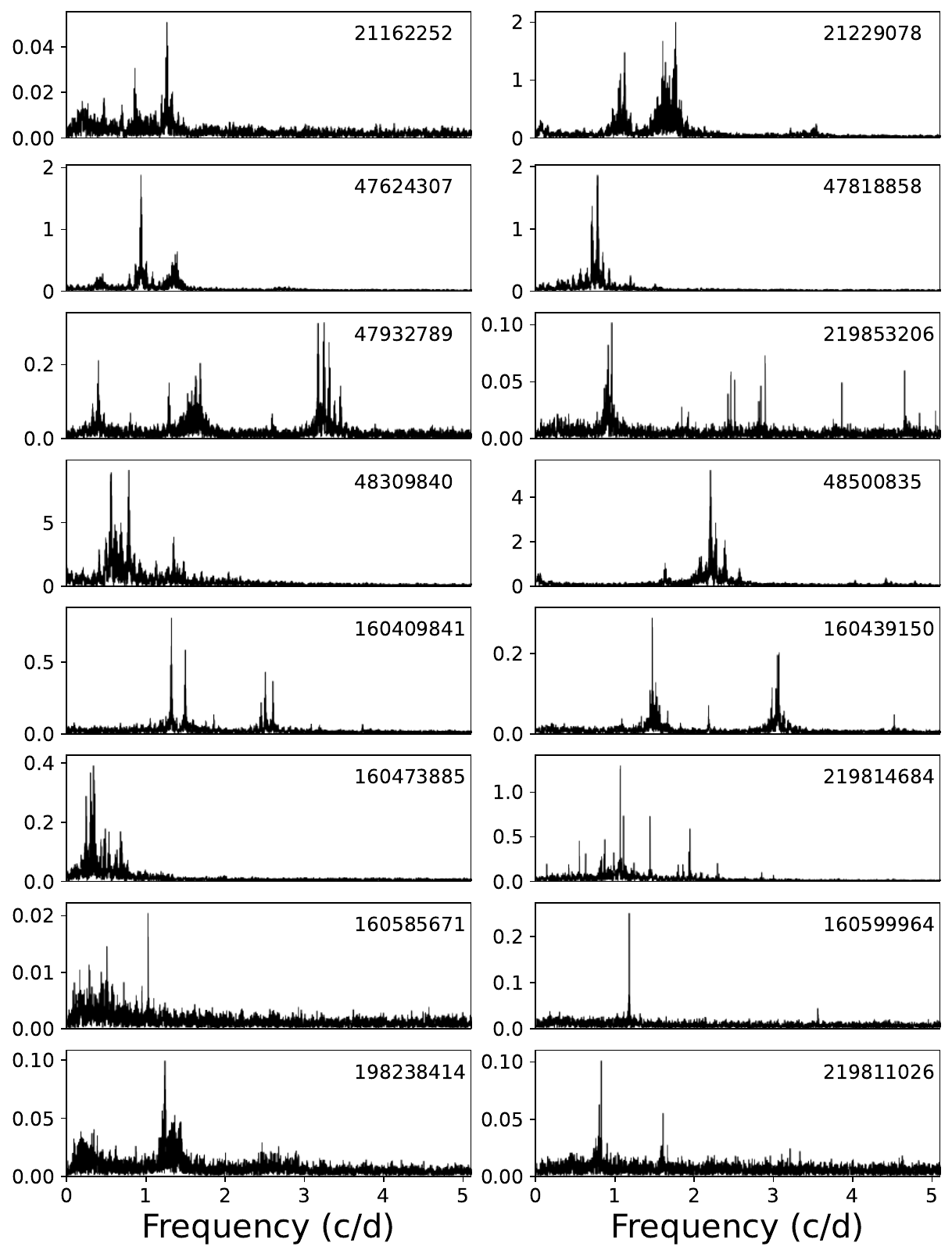}
\caption{Typical frequency spectra of GDOR stars. The four bottom-most plots show less certain cases. The scale on the vertical axis is in mmag. The labels show the TIC numbers.}
\label{Fig:GDORMosaic}
\end{figure*}

\begin{figure*}
\includegraphics[width=0.95\textwidth]{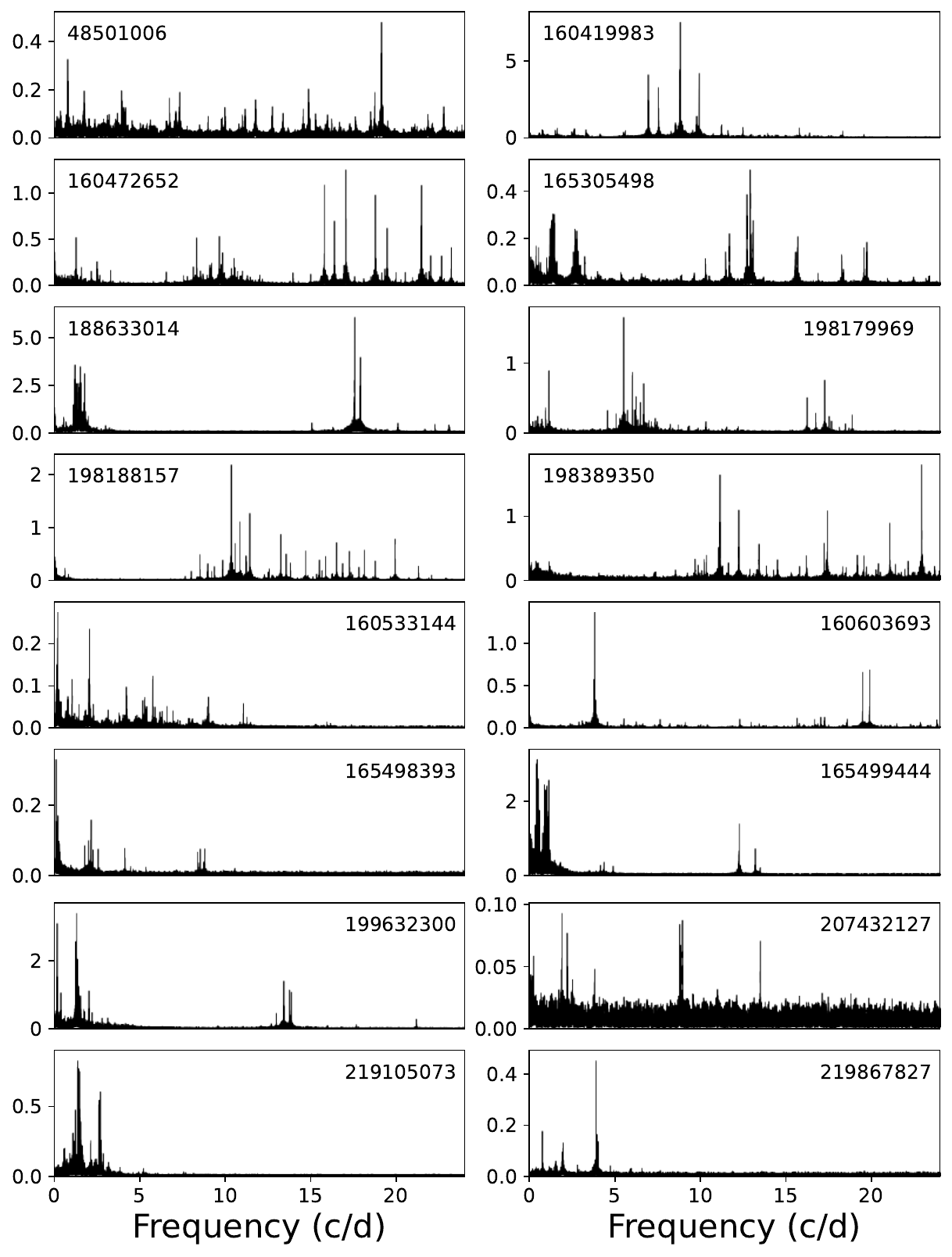}
\caption{Typical frequency spectra of GDOR and DSCT Hybrid stars. The eight upper panels show DSCT dominant, while the eight bottom panels show the GDOR dominant hybrids. TIC 198188157 and 198389350 show weak peaks in the GDOR regime and, thus, are less certain hybrids. TIC 219105073 and 219867827 show only low-amplitude peaks in the DSCT regime and, thus, are less certain hybrids. The less certain hybrids can possibly be classified as pure GDOR or DSCT stars. The scale on the vertical axis is in mmag.}
\label{Fig:HYBSosaic}
\end{figure*}

\begin{figure*}
\includegraphics[width=0.95\textwidth]{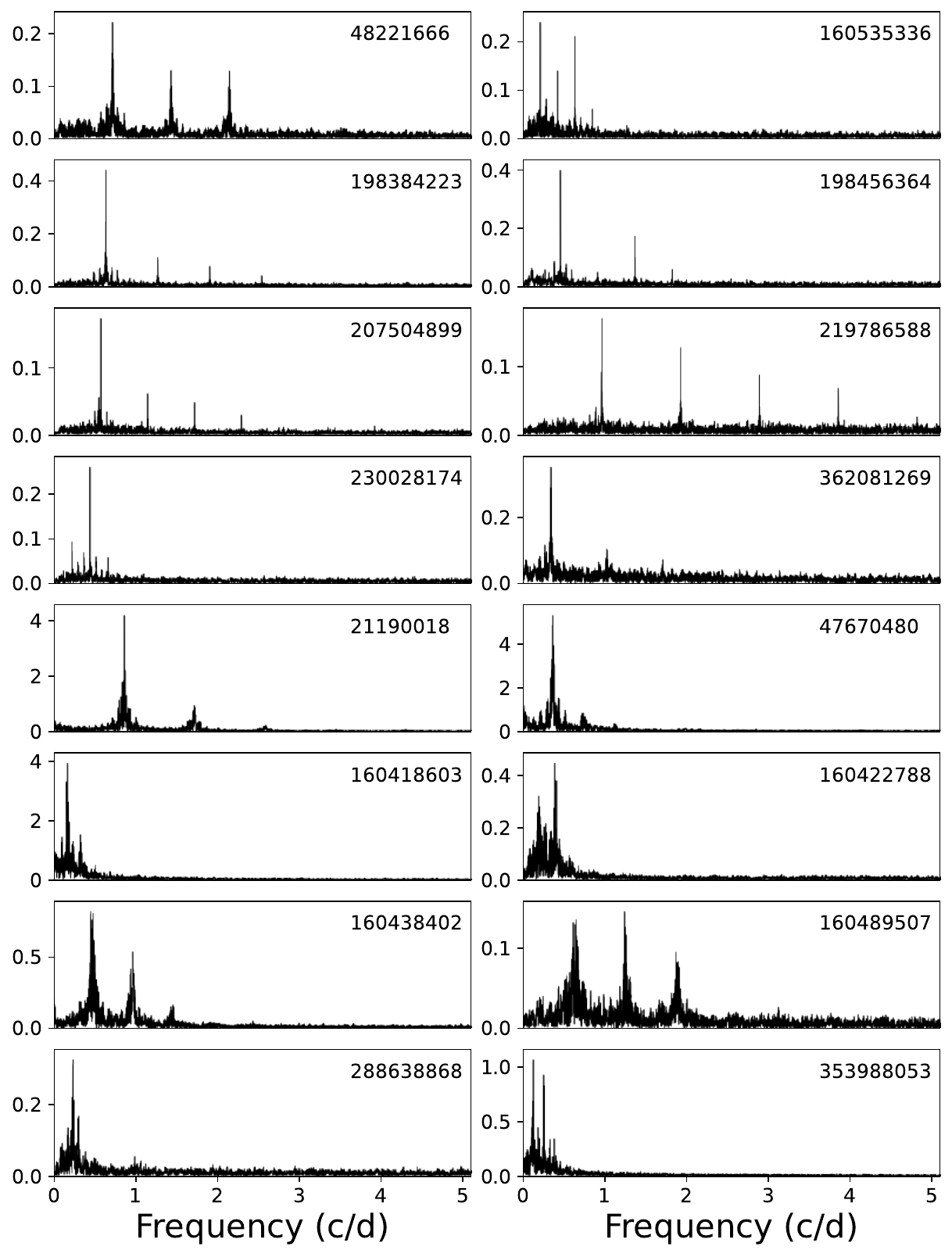}
\caption{Typical frequency spectra of ROT (eight upper panels) and ROTS stars (eight bottom panels). The scale on the vertical axis is in mmag.}
\label{Fig:ROTSMosaic}
\end{figure*}

\begin{figure*}
\includegraphics[width=0.95\textwidth]{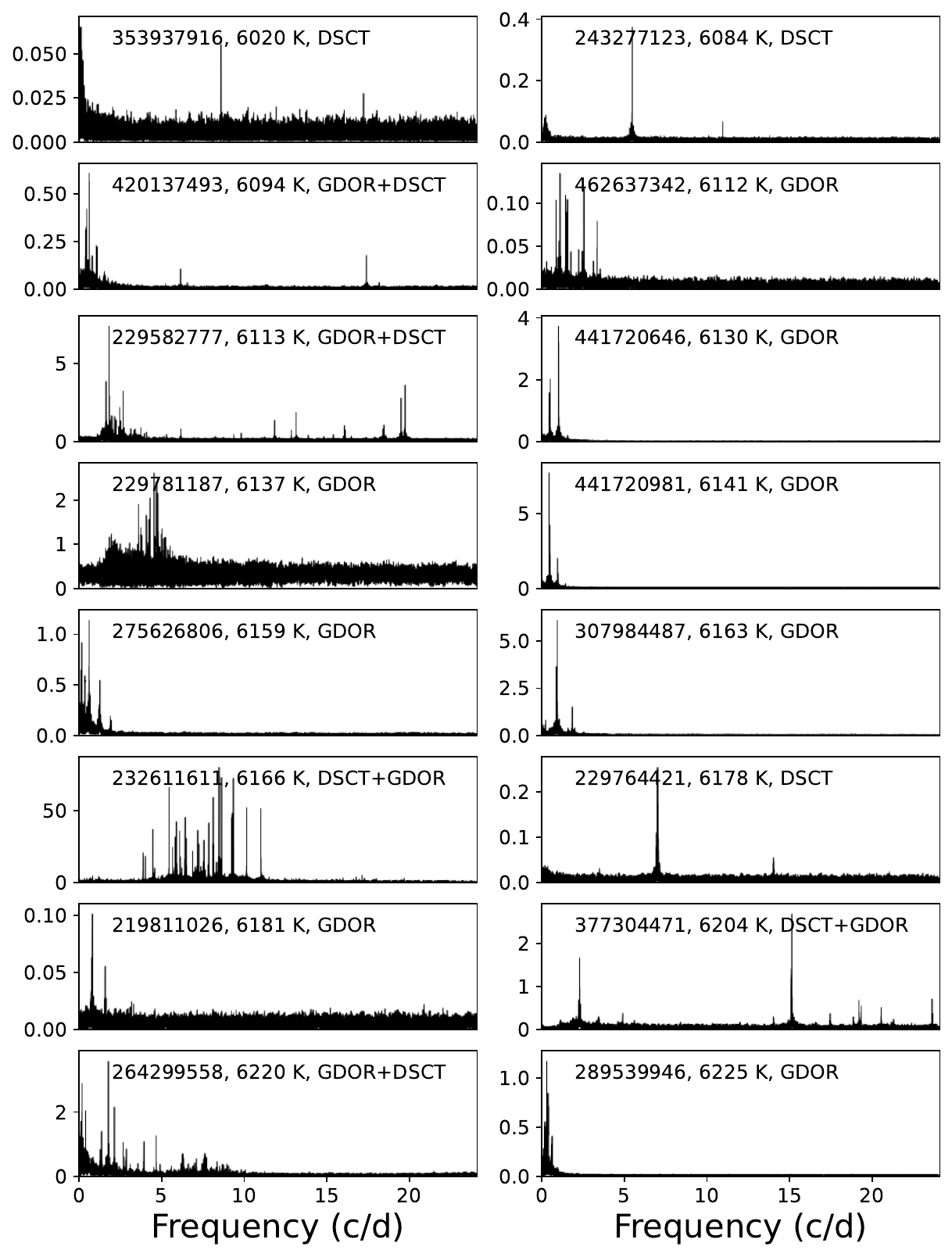}
\caption{Frequency spectra of the coolest pulsating stars in our sample. The first number is the TIC number, the second number gives the temperature from TIC catalogue. The scale on the vertical axis is in mmag.}
\label{Fig:Cool}
\end{figure*}

\end{appendix}

\end{document}